\documentstyle[12pt]{article}
\begin{document}
\title{\bf\Large
The algebraic Bethe ansatz for open $A_{2n}^{(2)}$ vertex model}

\author{
{\bf Guang-Liang Li$^{a}$ , Kang-Jie Shi$^{b}$ and Rui-Hong
Yue$^{b}$
} \\
{ \normalsize \it $^a$ Department of Applied Physics,
Xi'an Jiaotong University, Xi'an, 710049, P.R.China }\\
{\normalsize \sl$^b$ Institute of Modern Physics, Northwest
University, Xi'an, 710069, P.R.China }\\
{email address: leegl@mail.xjtu.edu.cn}}

\maketitle

 ABSTRACT:  We solve the $A_{2n}^{(2)}$ vertex model with all kinds of  diagonal reflecting
matrices by using the algebraic Behe ansatz, which includes
constructing the multi-particle states and achieving the
eigenvalue of the transfer matrix and corresponding Bethe ansatz
equations. When the model is $U_q(B_n)$ quantum invariant, our
conclusion agrees with that obtained by analytic Bethe ansatz
method.

\maketitle

\vspace{10mm}


KEYWORDS:  algebraic  Bethe ansatz; open boundary
\newpage
\tableofcontents
\section{Introduction}

\noindent

 In solving  integrable models, one of powerful
tools is the analytical Bethe ansatz method which was proposed by
Reshetikhin  for  close chains \cite{1}, and was generalized into
 quantum-algebra-invariant open chains \cite{2}.
 Then it was employed  to solve a family of both
quantum-algebra-invariant and non-quantum-algebra-invariant open
chains \cite{a2n2}-\cite{5}. For the result  is far from rigorous,
one more satisfactory approach would be the algebraic Bethe
ansatz(ABA), by which the Bethe ansatz equations and the
eigenstates can be obtained.

Recently,  the algebraic Bethe ansatz\cite{6}-\cite{8} has been
developed by Martins \cite{9}-\cite{11} for a large family of
vertex models with periodic boundary. Its generalizations  have
been applied into the vertex models with open boundary conditions
in Refs.\cite{12}-\cite{15} and Refs.\cite{ik}-\cite{osp12}. All
of these show that the ABA could  be suit  for the system with
higher rank algebra symmetry. Although some open boundary vertex
models, such as $A_{2n}^{(2)}$ model \cite{vv,mj}, have been
solved by the analytical Bethe ansatz or other methods, it is
still worthy to reconsider those by ABA. Here we will formulate
the algebraic Bethe ansatz solution for the $A_{2n}^{(2)}$ vertex
model with diagonal reflecting matrices.

The  $A_{2n}^{(2)}$ model at the case of $n=1$ is also called
Izergin-Korepin model \cite{19} which,  under the open boundary
conditions, can be related to the loop models \cite{5} and
flexible self- avoiding polymer chain \cite{20}. When $n>1$, the
model with trivial reflecting matrices was solved by the
analytical Bethe ansatz. However, for non-trivial  reflecting
matrix, the exact solutions remain unknown. In this paper, we
expect to solve the model with all kinds of  diagonal reflecting
matrices by using the algebraic Behe ansatz, which includes
constructing the multi-particle states and achieving the
eigenvalue of the transfer matrix and corresponding Bethe ansatz
equations. When the model is $U_q(B_n)$ quantum invariant, our
conclusion agrees with that obtained by analytic Bethe ansatz
method \cite{a2n2}.

The present paper is  organized  as following. In section 2 we
introduce $A_{2n}^{(2)}$ vertex model and list all diagonal
$K_{\pm }$ matrices governing  the  boundary terms in the
Hamiltonian. Section 3 devotes to construct m-particle
eigenfunctions and  to  derive out the eigenvalue and the Bethe
ansatz equations. A brief summary and discussion about  our main
result are included in section 4. Some necessary calculations and
coefficients are given as the Appendix.

\section{The vertex model and integrable boundary conditions}

The R matrix for  the $A_{2n}^{(2)}$ model used here is
\cite{a2n2}
 \begin{eqnarray}
   \hspace{-2mm}R^{(n)}(u) &=&a_n(u)\sum_{i\ne \bar{i}}E_{ii}\otimes
  E_{ii}+b_n(u)\sum_{i\ne j,\bar{j}}E_{ii}\otimes E_{jj}
   +\Big(\sum_{i<\bar{i}}c_n(u,i)+\sum_{i>\bar{i}}\bar{c}_n(u,i)\Big)
   E_{i\bar{i}}\otimes E_{\bar{i}i} \nonumber\\
   & &+
   \Big(\sum_{i<j,j\ne\bar{i}}d_n(u,i,j)+\sum_{i>j,j\ne\bar{i}}\bar{d}_n(u,i,j)\Big)
   E_{i{j}}\otimes E_{\bar{i}\bar{j}}
   +e_n(u)\sum_{i\ne \bar{i}}E_{ii}\otimes E_{\bar{i}\hspace{0.2mm} \bar{i}}
   \nonumber\\
   & &+f_n(u)E_{n+1n+1}\otimes E_{n+1n+1}+
   \Big(g_n(u)\sum_{i<j,j\ne\bar{i}}+\bar{g}_n(u)\sum_{i>j,j
   \ne\bar{i}}\Big)E_{ij}\otimes E_{{j}i} ,
 \label{rm}
 \end{eqnarray}
where
\begin{eqnarray}
a_n(u)&=&2\sinh(\frac{u}{2}-2\eta)\cosh(\frac{u}{2}-(2n+1)\eta),\nonumber\\
b_n(u)&=&2\sinh(\frac{u}{2})\cosh(\frac{u}{2}-(2n+1)\eta),\nonumber\\
c_n(u,i)&=&2e^{-u+2i\eta}\sinh((2i-(2n+1))\eta)\sinh(2\eta)-2e^{(2i-(2n+1))\eta}\sinh(2\eta)\cosh(2i\eta),\nonumber\\
\bar{c}_n(u,i)&=&2e^{u-2\bar{i}\eta}\sinh(((2n+1)-2\bar{i})\eta)\sinh(2\eta)-2e^{((2n+1)-2\bar{i})\eta}\sinh(2\eta)\cosh(2\bar{i}\eta),\nonumber\\
d_n(u,i,j)&=&-2e^{-\frac{u}{2}+(2n+1+2(\tilde{i}-\tilde{j}))\eta}\sinh(2\eta)\sinh(\frac{u}{2}),\nonumber\\
\bar{d}_n(u,i,j)&=&2e^{\frac{u}{2}+(2(\tilde{i}-\tilde{j})-2n-1)\eta}\sinh(2\eta)\sinh(\frac{u}{2}),\nonumber\\
e_n(u)&=&2\sinh(\frac{u}{2})\cosh(\frac{u}{2}-(2n-1)\eta),\nonumber\\
f_n(u)&=&b_n(u)-2\sinh(2\eta)\cosh((2n+1)\eta),\nonumber\\
g_n(u)&=&-2e^{-\frac{u}{2}}\sinh(2\eta)\cosh(\frac{u}{2}-(2n+1)\eta),\nonumber\\
\bar{g}_n(u)&=&-2e^{\frac{u}{2}}\sinh(2\eta)\cosh(\frac{u}{2}-(2n+1)\eta).
 \label{2}\\
i+\bar{i}&=&2n+2,\qquad \tilde{i}=\left\{\begin{array}{ll}
i+\frac{1}{2},& 1\le i< n+1\\
i,& i=n+1\\
i-\frac{1}{2},& n+1< i\le 2n+1
\end{array}\right.
\end{eqnarray}
  This  R matrix satisfies the following properties
\begin{eqnarray}
regularity&:&R^{(n)}_{12}(0)=\rho_n(0)^{\frac{1}{2}}{\cal P}_{12},\nonumber\\
unitarity&:&R^{(n)}_{12}(u)R^{(n)}_{21}(-u)=\rho_n(u),\nonumber\\
PT-symmetry&:&{\cal P}_{12}R^{(n)}_{12}(u){\cal P}_{12}=[R^{(n)}_{12}]^{t_1t_2}(u),\nonumber\\
crossing-unitarity&:&M_1R^{(n)}_{12}(u)^{t_2}M_1^{-1}R^{(n)}_{12}(-u-2\xi_n)^{t_1}=\rho_n(u+\xi_n),\nonumber
\end{eqnarray}
with $\rho_n(u)=a_n(u)a_n(-u),\ \xi_n=-\sqrt{-1}\pi-2(2n+1)\eta,\
M^i_{j}=\delta_{ij}e^{4(n+1-\tilde{i})\eta}$. The exchange
operator $\cal P$ is given by ${\cal
P}^{ij}_{kl}=\delta_{il}\delta_{jk}$, and $t_i$ denotes the
transposition in  $i$-th space. It also satisfies the Yang-Baxter
equation(YBE)\cite{6}
\begin{eqnarray}
R^{(n)}_{12}(u-v)R^{(n)}_{13}(u)R^{(n)}_{23}(v)=R^{(n)}_{23}(v)R^{(n)}_{13}(u)R^{(n)}_{12}(u-v),
\label{ybe}
\end{eqnarray}
$R^{(n)}_{12}(u)=R^{(n)}(u)\otimes 1, R^{(n)}_{23}(u)=1\otimes
R^{(n)}(u)$ etc. $R^{(n)}_{21}={\cal P}_{12}R^{(n)}_{12}{\cal
P}_{12}$. Define  the following reflection equations
\begin{equation}
 R^{(n)}_{12}(u-v)\stackrel{1}{K}_{-}(u)R^{(n)}_{21}(u+v) \stackrel{2}{K}_{-}(v)=
 \stackrel{2}{K}_{-}(v)R^{(n)}_{12}(u+v)\stackrel{1}{K}_{-}(u)R^{(n)}_{21}(u-v),
 \label{r1}
 \end{equation}
\begin{eqnarray}
 & &R^{(n)}_{12}(-u+v)\stackrel{1}{K}^{t_1}_{+}(u)\stackrel{1}{M^{-1}}R^{(n)}_{21}
 (-u-v-2\xi_n)\stackrel{1}{M}\stackrel{2}{K}^{t_2}_{+}(v)\nonumber\\
&=&\stackrel{2}{K}^{t_2}_{+}(v)\stackrel{1}{M}R^{(n)}_{12}(-u-v-2\xi_n)\stackrel{1}{M^{-1}}
\stackrel{1}{K}^{t_1}_{+}(u)R^{(n)}_{21}(-u+v),
 \label{r2}
 \end{eqnarray}
where $\stackrel{1}{K}_{\pm}(u)=K_{\pm}(u)\otimes
1,\stackrel{2}{K}_{\pm}(u)=1\otimes K_{\pm}(u)$. Then the transfer
matrix  defined as
\begin{equation}
t(u)={\rm tr}K_{+}(u)U(u) \label{tru}
\end{equation}
constitutes an one-parameter commutative family, i.e.
$[t(u),t(v)]=0$. Here
\begin{eqnarray}
U(u)&=&T(u)K_{-}(u)T^{-1}(-u),\label{Uform}\\
T(u)&=&R^{(n)}_{01}(u)R^{(n)}_{02}\cdots
R^{(n)}_{0N}(u).\label{Tform}
\end{eqnarray}
 The
corresponding integrable open chain Hamiltonian takes the form
\begin{equation}
H=\sum^{N-1}_{k=1}H_{k,k+1}
+\frac{1}{2}\stackrel{1}{K'}_{-}(0)+\frac{{\rm
tr}\stackrel{0}{K_{+}}(0)H_{N,0}}{{\rm tr}K_{+}(0)}, \label{hh}
\end{equation}
with $H_{k,k+1}={\cal P}_{k,k+1}R_{kk+1}'(u)|_{u=0}$.
 The general solutions of Eq.(\ref{r1}) have been
obtained in Ref.\cite{A2n}. The trivial and nontrivial diagonal
reflecting matrices take the form
\begin{eqnarray}
K^{(1)}_{-}(u,n)_i
& =& 1, \qquad i=1,2,\cdots, 2n+1 \label{kn1}\\[3mm]
K^{(2)}_{-}(u,n,p_-)_{i}
& =&\left\{\begin{array}{l}
           e^{-u}\Big[c_-\cosh(\eta)+\sinh(u-2(2p_--n)\eta)\Big],
                 \quad(1\le i\le p_-)\\[2mm]
           c_-\cosh(u+\eta)-\sinh(2(2p_--n)\eta),
                \quad(p_-+1\le i\le 2n+1-p_-)\\[2mm]
           e^{u}\Big[c_-\cosh(\eta)+\sinh(u-2(2p_--n)\eta)\Big].
                 \quad(2n+p_-+2\le i\le 2n+1)\end{array}\right.
                 \label{kn2}\\[5mm]
K^{(1)}_{+}(u,n)_i
&=&e^{4(n+1-\tilde{i})\eta}, \qquad i=1,2,\cdots, 2n+1
\label{kp1}\\[3mm]
K^{(2)}_{+}(u,n,p_+)_{i}
& =&\left\{\begin{array}{l}
      e^{4(n+1-\tilde{i})\eta+u-2(2n+1)\eta}
           \Big[c_+\cosh(\eta)+\sinh(u-2(3n-2p_++1)\eta)\Big],\\
           \qquad\qquad\qquad\qquad(1\le i\le p_+)\\[2mm]
      e^{4(n+1-\tilde{i})\eta}\Big[c_+\cosh(u-(4n+3)\eta)
      +\sinh(2(2p_+-n)\eta)\Big],\\
         \qquad\qquad\qquad\qquad(p_++1\le i\le 2n+1-p_+)\\[2mm]
      e^{4(n+1-\tilde{i})\eta-u+2(2n+1)\eta}
      \Big[c_+\cosh(\eta)+\sinh(u-2(3n-2p_++1)\eta)\Big],\\
          \qquad\qquad\qquad\qquad (2n+p_++2\le i\le 2n+1)\end{array}\right.\label{kp2}
\end{eqnarray}
where $c_-^2=c_+^2=-1$, $p_{\pm}$ are integer numbers varying from
$1$ to $n$.


\section{The algebraical  Bethe ansatz}

In this section, we will present the main procedure of solving
open  $A_{2n}^{(2)}$ model by using the nested Bethe ansatz. We
begin with introducing the vacuum state.

\subsection{The vacuum state}
Firstly,  we write the double-monodromy matrix (\ref{Uform}) as
\begin{equation}
U(u)=\left(\begin{array}{cccccc}
 A(u)& B_1(u)& B_2(u)& \cdots & B_{2n-1}&F(u)\\
D_1(u)& A_{11}(u)& A_{12}(u)&\cdots &A_{12n-1}(u) &E_1(u)\\
D_2(u)& A_{21}(u)& A_{22}(u)&\cdots &A_{22n-1}(u) &E_2(u)\\
\vdots&\vdots&      \vdots&  \vdots& \vdots&\vdots \\
D_{2n-1}(u)& A_{2n-11}(u)& A_{2n-12}(u)&\cdots &A_{2n-12n-1}(u) &E_{2n-1}(u)\\
 G(u)& C_1(u)& C_2(u)&\cdots &C_{2n-1}(u) & A_2(u)
\end{array}\right). \label{uu}
\end{equation}
With the help of Eqs.(\ref{ybe},\ref{r1}), we can prove that
$U(u)$ in eq.(\ref{uu})  satisfy the following equation
\begin{equation}
 R^{(n)}_{12}(u-v)\stackrel{1}U(u)R^{(n)}_{21}(u+v) \stackrel{2}U(v)=
 \stackrel{2}U(v)R^{(n)}_{12}(u+v)\stackrel{1}U(u)R^{(n)}_{21}(u-v).
 \label{dbmzj}
 \end{equation}
Applying  the double-row monodromy matrix eq.(\ref{uu}) on the
vacuum state $|0\rangle=\prod^{\otimes N} (1,0,\cdots,0)^t$, we
can find
\begin{eqnarray}
& &D_a(u)|0\rangle=0,\hspace{4mm}C_a(u)|0\rangle=0,
\hspace{4mm}G(u)|0\rangle=0,\nonumber \\
& &B_a(u)|0\rangle\ne 0, \hspace{4mm}E_a(u)|0\rangle\ne 0,
\hspace{4mm} F(u)|0\rangle\ne 0,\nonumber \\
& &A_{aa}(u)|0\rangle\ne 0,\hspace{3mm}A_{ab}(u)|0\rangle= 0\ \
(a\ne b
),\nonumber\\
& &A(u)|0\rangle\ne 0,\hspace{5mm}A_{2}(u)|0\rangle\ne
0.\hspace{4mm} (a=1,2,\cdots,2n-1)\label{avc}
\end{eqnarray}
From Eq.(\ref{avc}), we can see that $D_a,C_a$  and $B_a, E_a, F$
play the role of annihilation operators and creation operators on
the vacuum state, respectively. The $A, A_{aa}, A_2$ are diagonal
operators on the vacuum state. Introducing two new operators (a
not very long calculation is omitted here)
\begin{eqnarray}
\tilde{A}_{ab}(u)&=&A_{ab}(u)-\tilde{f}_1(u)A(u)\delta_{ab},\label{d1}\\
\tilde{A}_2(u)&=&A_2(u)-\tilde{f}_3(u)A(u)-\tilde{f}_2(u)\sum_{a=1}^{2n-1}e^{4(n-\tilde{a})\eta}\tilde{A}_{aa}(u),\label{d2}
\end{eqnarray}
where
\begin{eqnarray}
& & \tilde{f}_1(u)=\frac{\bar{g}_n(2u)}{a_n(2u)},
\qquad\tilde{f}_3(u)=\frac{\bar{c}_{n}(2u,2n+1)}{a_n(2u)},\qquad
\tilde{f}_2(u)=-\frac{e^{u-4\eta}\sinh(2\eta)}{\sinh(u-4n\eta)},\nonumber\\
 && \tilde{a}=\left\{\begin{array}{ll}
a+\frac{1}{2},& 1\le a< n\\
a,& a=n\\
a-\frac{1}{2},& n+1< a\le 2n-1
\end{array}\right.
\end{eqnarray}
  we have
\begin{eqnarray}
&&A(u)|0\rangle=K_-(u)_{1}{[a_n(u)]^{2N}}{\rho_n(u)^{-N}}
|0\rangle=\omega_1 (u)|0\rangle,\\[3mm]
&&\tilde{A}_{aa}(u)|0\rangle=(K_-(u)_{a+1}-\tilde{f}_1(u)
K_-(u)_{1}){[b_n(u)]^{2N}}{\rho_n(u)^{-N}}|0\rangle
=k^-({u})_a\omega(u)|0\rangle,\\[3mm]
&&\tilde{A}_2(u)|0\rangle=\Big\{K_-(u)_{2n+1}-\tilde{f}_2(u)
\sum_{a=1}^{2n-1}e^{4(n-\tilde{a})\eta}\Big(K_-(u)_{a+1}
-\tilde{f}_1(u)K_-(u)_{1}\Big)\nonumber\\
&&\hspace{4cm}-\tilde{f}_3(u)K_-(u)_{1}\Big\}
{[e_n](u)^{2N}}{\rho_n(u)^{-N}}|0\rangle=\omega_{2n+1}(u)|0\rangle.
\end{eqnarray}
\noindent In terms of new operators, the transfer matrix
(\ref{tru}) can be rewritten as
\begin{eqnarray}
&&t(u)={w}_1(u)A(u)+\sum_{a=1}^{2n-1}{w}(u)k^+_{a}(u)\tilde{A}_{aa}(u)
+{w}_{2n+1}(u)\tilde{A}_2(u)\label{tru1}
\end{eqnarray}
with
\begin{eqnarray}
&&{w}_1(u)=K_+(u)_1+\tilde{f}_3(u)K_+(u)_{2n+1}+\tilde{f}_1(u)
\sum_{a=1}^{2n-1}K_+(u)_{a+1},\nonumber\\
&&{w}(u)k^+_{a}(u)=K_+(u)_{a+1}+e^{4(n-\tilde{a})
\eta}\tilde{f}_2(u)K_+(u)_{2n+1},\hspace{4mm}
{w}_{2n+1}(u)=K_+(u)_{2n+1}.
\end{eqnarray}
The explicit expression of  coefficient functions $\omega$'s and
$w$'s can be seen at the case $j=0$ in Appendix B,
$k^{\mp}({u})=K^{(1)}_{\mp}(\tilde{u},n-1)$ or
$K^{(2)}_{\mp}(\tilde{u},n-1,p_{\mp}-1)$ depend on the choice of
boundary, $\tilde{u}=u-2\eta$.

\subsection{The Fundamental commutation relations}

 In order to construct the general m-particle state, we need to find the commutation relations between the creation,
 diagonal and annihilation fields. Here we only provide  some important commutation relations.
 Taking some components of eq.(\ref{dbmzj}), we can obtain the
following fundamental commutation relations
\begin{eqnarray}
 & & B_a(u)B_b(v)+\delta_{\bar{a}b} g_1(u,v,a)F(u)A(v)
 + g_2(u,v,a)F(u)\tilde{A}_{\bar{a}b}(v)\nonumber\\
& &=\hat{r}(u_-)^{dc}_{ba} [B_d(v)B_c(u)+\delta_{\bar{d}c}
g_1(v,u,d)F(v)A(u)
 + g_2(v,u,d)F(v)\tilde{A}_{\bar{d}c}(u)],
\label{b1b1}
\end{eqnarray}
\begin{eqnarray}
A(u)B_a(v)&=&a^1_1(u,v)B_a(v)A(u)+a^1_2(u,v)B_a(u)A(v)+a^1_3(u,v)B_d(u)\tilde{A}_{da}(v)\nonumber\\
 & &+ a^1_4(u,v,\bar{a})F(u)D_{\bar{a}}(v)+a^1_5(u,v)F(u)C_a(v)+ a^1_6(u,v,\bar{a})F(v)D_{\bar{a}}(u),\label{ab1}\\[3mm]
\tilde{A}_{ab}(u)B_c(v)&=&{\tilde{r}(u_+)^{ae}_{dg}\bar{r}(u_-)^{gf}_{cb}}B_e(v)\tilde{A}_{df}(u)
+R^{A}_1(u,v)^{af}_{cb}B_f(u)A(v)\nonumber\\
& & +R^{A}_2(u,v)^{af}_{db}B_f(u)\tilde{A}_{dc}(v)
 +\delta_{\bar{b}c} R^{A}_3(u,v,\bar{b})E_a(u)A(v)\nonumber\\
 & & + R^{A}_4(u,v,\bar{b})E_a(u)\tilde{A}_{\bar{b}c}(v)+  R^{A}_5(u,v)^{af}_{cb}F(u)D_{\bar{f}}(v)\nonumber\\
 & &+\delta_{ab}R^{A}_6(u,v)F(u)C_c(v)+ R^{A}_7(u,v)^{af}_{cb}F(v)D_{\bar{f}}(u)\nonumber\\
 & & +R^{A}_8(u,v)^{af}_{cb}F(v)C_f(u),\label{d1b1}\\
\tilde{A}_2(u)B_a(v)&=&a^3_1(u,v)B_a(v)\tilde{A}_2(u)+a^3_2(u,v)B_a(u)A(v)+a^3_3(u,v)B_d(u)\tilde{A}_{da}(v)\nonumber\\
 & &+ a^3_4(u,v,{a})E_{\bar{a}}(u)A(v)+ a^3_5(u,v,\bar{d})E_d(u)\tilde{A}_{\bar{d}a}(v)
 +  a^3_6(u,v,\bar{a})F(u)D_{\bar{a}}(v)\nonumber\\
 & &+a^3_7(u,v)F(u)C_a(v)+
 a^3_8(u,v,\bar{a})F(v)D_{\bar{a}}(u)+a^3_{9}(u,v)F(v)C_a(u),\label{d2b1}\\
A(u)F(v)&=& b^1_1(u,v)F(v)A(u)+b^1_2(u,v)F(u)A(v)+b^1_3(u,v,d)F(u)\tilde{A}_{dd}(v)\nonumber\\
 & &+b^1_4(u,v)F(u)\tilde{A}_2(v)+ b^1_5(u,v,d)B_{\bar{d}}(u)B_d(v)+b^1_6(u,v)B_d(u)E_d(v),\\
\tilde{A}_{ab}(u)F(v)&=&b^2_1(u,v)F(v)\tilde{A}_{ab}(u)+\delta_{ab}b^2_2(u,v)F(u)A(v)
+R^{F}_1(u,v)^{dc}_{b\bar{a}}F(u)\tilde{A}_{\bar{d}c}(v)\nonumber\\
& &+\delta_{ab}b^2_3(u,v)F(u)\tilde{A}_2(v)+
R^F_2(u,v)^{dc}_{b\bar{a}}B_d(u)B_c(v)
+R^F_3(u,v)^{ac}_{db}B_c(u)E_d(v)\nonumber\\
 & &+b^2_4(u,v)E_a(u)B_b(v)+ b^2_5(u,v,\bar{b})E_a(u)E_{\bar{b}}(v),\\[2mm]
\tilde{A}_2(u)F(v)&=&b^3_1(u,v)F(v)\tilde{A}_2(u)+b^3_2(u,v)F(u)A(v)+b^3_3(u,v,d)F(u)\tilde{A}_{dd}(v)\nonumber\\
& &+b^3_4(u,v)F(u)\tilde{A}_2(v)+ b^3_5(u,v,\bar{d})B_{\bar{d}}(u)B_d(v)+b^3_6(u,v)B_d(u)E_d(v)\nonumber\\
& &+b^3_7(u,v)E_d(u)B_d(v)+ b^3_8(u,v,d)E_d(u)E_{\bar{d}}(v).
\end{eqnarray}
Besides the above fundamental commutation relations, we also need
the following necessary commutation relations
\begin{eqnarray}
D_a(u)B_b(v)&=& R^D_1(u,v)^{ac}_{db}B_c(v)D_d(u)+
c^1_1(u,v,\bar{a})B_{\bar{a}}(v)C_b(u)
+\delta_{ab}c^1_2(u,v)F(v)G(u)\nonumber \\
& & + c^1_3(u,v,\bar{a})B_{\bar{a}}(u)C_b(v)+c^1_4(u,v)E_a(u)C_b(v)+\delta_{ab}c^1_5(u,v)A(v)A(u)\nonumber \\
& & +c^1_6(u,v)A(v)\tilde{A}_{ab}(u)+\delta_{ab}c^1_7(u,v)A(u)A(v)+c^1_8(u,v)A(u)\tilde{A}_{ab}(v)\nonumber \\
& & +c^1_{9}(u,v)\tilde{A}_{ab}(u)A(v)+c^1_{10}(u,v)\tilde{A}_{ad}(u)\tilde{A}_{db}(v),\label{c1b1}\\
C_a(u)B_b(v)&=& R^C_1(u,v)^{dc}_{ba}B_d(v)C_c(u)+
R^C_2(u,v)^{\bar{a}c}_{db}B_c(v)D_d(u)+\delta_{a\bar{b}} c^2_1(u,v,\bar{a})F(v)G(u)\nonumber \\
& &
+c^2_2(u,v)B_a(u)C_b(v)+c^2_3(u,v,\bar{a})E_{\bar{a}}(u)C_b(v)+\delta_{a\bar{b}}
c^2_4(u,v,\bar{a})A(v)A(u) \nonumber \\
 & &+
R^C_3(u,v)^{dc}_{ba}A(v)\tilde{A}_{\bar{d}c}(u)
+\delta_{a\bar{b}} c^2_5(u,v,\bar{a})A(v)\tilde{A}_2(u)\nonumber \\
& & +\delta_{a\bar{b}} c^2_6(u,v,\bar{a})A(u)A(v)+
c^2_{7}(u,v,\bar{a})A(u)\tilde{A}_{\bar{a}b}(v)\nonumber \\
& & + R^C_4(u,v)^{dc}_{ba}\tilde{A}_{\bar{d}c}(u)A(v)+
R^C_5(u,v)^{dc}_{ea}\tilde{A}_{\bar{d}c}(u)\tilde{A}_{eb}(v)\nonumber \\
& &+\delta_{a\bar{b}}c^2_{8}(u,v,\bar{a})\tilde{A}_2(u)A(v)+
c^2_{9}(u,v,\bar{a})\tilde{A}_2(u)\tilde{A}_{\bar{a}b}(v),\label{c2b1}\\
B_a(u)E_b(v)&=& R^{be}_1(u,v)^{ca}_{bd}E_c(v)B_d(u)+
R^{be}_2(u,v)^{dc}_{\bar{b}a}B_d(v)B_c(u)
+\delta_{ab}e^1_1(u,v,a)F(v)A(u)\nonumber \\
& &+
R^{be}_3(u,v)^{dc}_{\bar{b}a}F(v)\tilde{A}_{\bar{d}c}(u)+\delta_{ab}e^1_2(u,v,a)F(u)A(v)
+R^{be}_4(u,v)^{ca}_{bd}F(u)\tilde{A}_{cd}(v)\nonumber \\
& & +\delta_{ab}e^1_3(u,v,a)F(u)\tilde{A}_2(v),\label{b1b2}
\end{eqnarray}
where all the repeated indices sum over 1 to $2n-1$, $u_{\pm}=u\pm
v$  and
\begin{eqnarray}
&&g_1(u,v,a)
=-\frac{d_n(u_-,1,\bar{a})b_n(2v)}{e_n(u_-)a_n(2v)},\hspace{4mm}
g_2(u,v,a) =\frac{d_n(u_+,1,\bar{a})}{b_n(u_+)}\, \qquad
a+\bar{a}=2n\ .
\end{eqnarray}
\noindent The $\hat{r}(u),\tilde{r}(u)$ and $\bar{r}(u)$  are
given by
\begin{eqnarray}
&&\hat{r}(u)=\frac{1}{e_n(u)}\frac{b_n(u)}{a_n(u)}R^{(n-1)}(u),\hspace{2mm}\tilde{r}(u)
=\frac{1}{a_n(u)}R^{(n-1)}(u-4\eta),\hspace{2mm}\bar{r}(u)=\frac{1}{e_n(u)}R^{(n-1)}(u),\nonumber
\end{eqnarray}
respectively. The other coefficients are not presented here for
their long and tedious expressions.

\subsection{The m-particle state}

 \noindent Inferred from the commutation relation Eq.(\ref{b1b1}),
 we can construct the general m-particle state as follow. Let
\begin{eqnarray}
& & \hspace{-6mm}\Phi_m^{b_1\cdots b_m}(v_1,\cdots,v_m)= B_{b_1}(v_1)\Phi_{m-1}^{b_2\cdots b_m}(v_2,\cdots,v_m)\nonumber \\
& & +F(v_1)\sum_{i=2}^m \Phi_{m-2}^{d_3\cdots
d_m}(v_2,\cdots,\check{v}_i,\cdots,v_m) S^{d_2\cdots
d_m}_{b_2\cdots b_m}(v_i;\{\check{v}_1,\check{v}_i\})\nonumber \\
& & \hspace{3mm}\times \Lambda_1^{m-2}(v_i;
\{\check{v}_1,\check{v}_i\})g_{1}(v_1,v_i,b_1)A(v_i)
\delta_{\bar{b}_1d_2}\nonumber \\
& & +F(v_1)\sum_{i=2}^m \Phi_{m-2}^{d_3\cdots
d_m}(v_2,\cdots,\check{v}_i,\cdots,v_m)
[\tilde{T}^{m-2}(v_i;\{\check{v}_1,\check{v}_i\})^{d_3\cdots
d_m}_{c_3\cdots c_m}]_{\bar{b}_1c_2}\nonumber \\
& & \hspace{3mm}\times S^{c_2\cdots c_m}_{b_2\cdots
b_m}(v_i;\{\check{v}_1,\check{v}_i\})g_2(v_1,v_i,b_1),
\label{nps2}
\end{eqnarray}
\noindent where
\begin{eqnarray}
S^{d_1\cdots d_m}_{b_1\cdots b_m}(v_i;\{\check{v}_i\})&=&
\hat{r}^{d_1d_2}_{c_2b_1}(v_1-v_i)\hat{r}^{c_2d_3}_{c_3b_2}(v_2-v_i)\cdots
\hat{r}^{c_{i-1}d_i}_{b_ib_{i-1}}(v_{i-1}-v_i)\prod^m_{j=i+1}\delta_{d_jb_j}
\nonumber \\[5mm] [\tilde{T}^{m}(u;\{{v}_m\})^{d_1\cdots
d_m}_{c_1\cdots c_m}]_{ab}&= &
\tilde{r}^{ad_1}_{h_1g_1}(u+v_1)\tilde{r}^{h_1d_2}_{h_2g_2}(u+v_2)\cdots
\tilde{r}^{h_{m-1}d_m}_{h_mg_m}(u+v_m)\tilde{A}_{h_mf_m}(u)\nonumber\\
& &
\bar{r}^{g_mf_m}_{c_mf_{m-1}}(u-v_m)\bar{r}^{g_{m-1}f_{m-1}}_{c_{m-1}f_{m-2}}(u-v_{m-1})
\cdots \bar{r}^{g_1f_1}_{c_1b}(u-v_1)
\end{eqnarray}
with  $ S^{d_1\cdots d_m}_{b_1\cdots
b_m}(v_1;v_2,\cdots,v_m)=\prod_{i=1}^{m}\delta_{d_ib_i}$,
$[\tilde{T}^{0}(u)]_{ab}=\tilde{A}_{ab}(u)$,
$\Lambda_l^m(u;v_1,v_2,\cdots,v_m)= \prod_{i=1}^m a^l_1(u,v_i)$, $
(l=1,3)$, $\Phi_0=1, \Phi_1^{b_1}(v_1)=B_{b_1}(v_1)$. The
$\check{v}_i$ means missing of ${v}_i$ in the sequence.

Then the general m-particle state is defined by
\begin{eqnarray}
|\Upsilon_m(v_1,\cdots,v_m)\rangle=\Phi_m^{b_1\cdots
b_m}(v_1,\cdots,v_m)F^{b_1\cdots b_m}|0\rangle, \label{nps1}
\end{eqnarray}
which enjoys  the property
\begin{eqnarray}
& & \hspace{-6mm}\Phi_m^{b_1\cdots b_ib_{i+1}\cdots
b_m}(v_1,\cdots,v_i,v_{i+1},\cdots,v_m)F^{b_1\cdots b_m}|0\rangle= \nonumber\\
& & \Phi_m^{b_1\cdots a_ia_{i+1}\cdots
b_m}(v_1,\cdots,v_{i+1},v_{i},\cdots,v_m)\hat{r}^{a_{i+1}a_i}_{b_ib_{i+1}}(v_i-v_{i+1})F^{b_1\cdots
b_m}|0\rangle\ . \label{npsp}
\end{eqnarray}
It is easy to verify  Eq.(\ref{npsp}) excepting  $i=1$. But, the
proof for the case $i=1$  becomes very involved and are  omitted
here.


\subsection{The eigenvalue and Bethe equations}

We can apply  the operators $A$ 's on the eigenstate ansatz  and
obtain (see Appendix A)

\begin{eqnarray}
& &
\hspace{-10mm}{x}(u)|\Upsilon_m(v_1,\cdots,v_m)\rangle=|\tilde{\Psi}_{x}(u,\{v_m\})\rangle\nonumber\\
 & &+\sum_{i=1}^m h_1^x(u,v_i,d_1)|\Psi_{m-1}^{(1)}(u,v_i;\{v_m\})_{d_1d_1}\rangle\nonumber\\
& & +\sum_{i=1}^m  h_2^x(u,v_i,d)|\tilde{\Psi}_{m-1}^{(2)}(u,v_i;\{v_m\})_{dd}\rangle\nonumber\\
& &+\sum_{i=1}^m h_3^x(u,v_i,\bar{\alpha}_x)|\Psi_{m-1}^{(3)}(u,v_i;\{v_m\})_{\alpha_x\alpha_x}\rangle\nonumber\\
& &+\sum_{i=1}^m h_4^x(u,v_i,\bar{\alpha}_x)|\tilde{\Psi}_{m-1}^{(4)}(u,v_i;\{v_m\})_{\alpha_x\alpha_x}\rangle\nonumber\\
& & +\sum^{m-1}_{i=1}\sum_{j=i+1}^m
\tilde{H}^{A_{aa}}_{1,d_1}(u,v_i,v_j)|\tilde{\Psi}_{m-2}^{(5)}(u,v_i,v_j;\{v_m\})_{d_1}\rangle\nonumber\\
& & +\sum^{m-1}_{i=1}\sum_{j=i+1}^m
\tilde{H}^{A_{aa}}_{2,d_1}(u,v_i,v_j)|\tilde{\Psi}_{m-2}^{(6)}(u,v_i,v_j;\{v_m\})_{d_1}\rangle\nonumber\\
& & +\sum^{n-1}_{i=1}\sum_{j=i+1}^m
\tilde{H}^{A_{aa}}_{3,d_1}(u,v_i,v_j)|\tilde{\Psi}_{m-2}^{(7)}(u,v_i,v_j;\{v_m\})_{d_1}\rangle\nonumber\\
& & +\sum^{m-1}_{i=1}\sum_{j=i+1}^m
\tilde{H}^{A_{aa}}_{4,d_1}(u,v_i,v_j)|\tilde{\Psi}_{m-2}^{(8)}(u,v_i,v_j;\{v_m\})_{d_1}\rangle,
\label{nd1pns}
\end{eqnarray}
where  the expression of $|\tilde{\Psi}\rangle$'s and coefficients
$\tilde{H}^{x}_{j,d_1}$ ($j=1,2,3,4$) are  given in Appendix A,
$x=A,\tilde{A}_{aa},\tilde{A}_2$.
 Using  eq.(\ref{tru1}), we then get
\begin{eqnarray}
& & \hspace{-15mm}t(u)|\Upsilon_m(v_1,\cdots,v_m)\rangle
=w_1(u)\omega_1(u)\Lambda_1^m(u;v_1,\cdots,v_m)|\Upsilon_m(v_1,\cdots,v_m)\rangle\nonumber\\
& &+ {w}(u)\omega(u)\Lambda_2^m(u;v_1,\cdots,v_m)\Phi_m^{d_1\cdots
d_m}(v_1,\cdots,v_m)\tau_1(\tilde{u};\{\tilde{v}_m\})^{d_1\cdots
d_m}_{b_1\cdots b_m} F^{b_1\cdots
b_m}|0\rangle\nonumber\\
&
&+w_{2n+1}(u)\omega_{2n+1}(u)\Lambda_3^m(u;v_1,\cdots,v_m)|\Upsilon_m(v_1,\cdots,v_m)\rangle+
u.t. , \label{tupn}
\end{eqnarray}
where $u.t.$ denotes the unwanted terms,
\begin{eqnarray}
& & \hspace{-10mm}\tau_1(\tilde{u};\{\tilde{v}_m\})^{d_1\cdots
d_m}_{c_1\cdots
c_m}=\nonumber\\
& &
k^+({u})_a{L}(\tilde{u},\tilde{v}_1)^{ad_1}_{h_1g_1}{L}(\tilde{u},\tilde{v}_2)^{h_1d_2}_{h_2g_2}\cdots
{L}(\tilde{u},\tilde{v}_m)^{h_{m-1}d_m}_{h_mg_m}k^-({u})_{h_m}\nonumber\\
& & \times {L^{-1}}(-\tilde{u},\tilde{v}_m)^{h_mg_m}_{f_{m-1}c_m}
{L^{-1}}(-\tilde{u},\tilde{v}_{m-1})^{f_{m-1}g_{m-1}}_{f_{m-2}c_{m-1}}
\cdots
{L^{-1}}(-\tilde{u},\tilde{v}_1)^{f_1g_1}_{ac_1}.\label{taud}
\end{eqnarray}
with  $\tilde{v_i}=v_i-2\eta$, and
\begin{eqnarray}
& &
L(\tilde{u},\tilde{v})^{ab}_{cd}=R^{(n-1)}(\tilde{u}+\tilde{v})^{ab}_{cd},\nonumber\\
& &
L^{-1}(-\tilde{u},\tilde{v})^{ab}_{cd}=\frac{R^{(n-1)}(\tilde{u}-\tilde{v})^{ba}_{dc}}{\rho_{n-1}(\tilde{u}-\tilde{v})}.
\end{eqnarray}
Thus, we get the conclusion that$
|\Upsilon_m(v_1,\cdots,v_m)\rangle$ is the eigenstate of $t(u)$,
i.e.
\begin{eqnarray} & &
\hspace{-15mm}t(u)|\Upsilon_m(v_1,\cdots,v_m)\rangle
=\{w_1(u)\omega_1(u)\Lambda_1^m(u;v_1,\cdots,v_m)\nonumber\\
& &+ {w}(u)\omega(u)\Lambda_2^m(u;v_1,\cdots,v_m)
\Gamma_{1}(\tilde{u};\{\tilde{v}_{m}\};\{{v}_{m_{1}}^{(1)}\})\nonumber\\
&
&+w_{2n+1}(u)\omega_{2n+1}(u)\Lambda_3^m(u;v_1,\cdots,v_m)\}|\Upsilon_m(v_1,\cdots,v_m)\rangle \nonumber\\
& & =\Gamma(u;\{{v}_m\})|\Upsilon_m(v_1,\cdots,v_m)\rangle,
\label{tupn1}
\end{eqnarray} if the parameters satisfy
\begin{equation}
\tau_1(\tilde{u};\{\tilde{v}_m\}) F^{b_1\cdots
b_m}=\Gamma_{1}(\tilde{u};\{\tilde{v}_{m}\};\{{v}_{m_{1}}^{(1)}\})F^{b_1\cdots
b_m}\ , \label{tau1}
\end{equation}
\begin{equation}
\Gamma_{1}(\tilde{v}_i;\{\tilde{v}_{m}\};\{{v}_{m_{1}}^{(1)}\})=-\rho^{-\frac{1}{2}
}_{n-1}(0)\frac{\omega_1(v_i)\Lambda_1^{m-1}(v_i;
\{\check{v}_i\})} {\omega(v_i)\Lambda_2^{m-1}(v_i;
\{\check{v}_i\})}\beta_1(v_i),\hspace{2mm}(i=1,\cdots,m)\
,\label{BA1}
\end{equation}
 where
\begin{equation}
\beta_1(v_i)=\displaystyle \left\{\begin{array}{cc} \displaystyle
-\frac{2e^{2\eta}\sinh(v_i)\sinh(v_i-4n\eta)\cosh(v_i-(2n-1)\eta)}{\sinh(v_i-2\eta)},&{\rm
for\ the\
eq.(\ref{kp1})}\\[2mm]\displaystyle
-\frac{2e^{v_i}\sinh(v_i)\sinh(v_i-4n\eta)}{\sinh(v_i-2\eta)}
[\sinh(2\eta)+c_+\cosh(v_i-(2n+1)\eta)]\nonumber\\
\hspace{4mm}\times[\cosh((4p_+-4n-1)\eta)-c_+\sinh(v_i-2(n+1)\eta)].&{\rm
for\  the\ eq.(\ref{kp2})}
\end{array}\right.\ .\label{BA2}
\end{equation}
All unwanted terms  cancel out  by the following three kinds of
identities
\begin{eqnarray}
&&\beta_1(v_i)=T^{(d_1)}(v_i)\frac{w_1(u)a^1_2(u,v_i)+\sum_{d=1}^{2n-1}w_{{d}+1}(u)R^A_1(u,v_i)^{dd_1}_{d_1d}
+w_{2n+1}(u)a^3_2(u,v_i)}
{w_1(u)a^1_3(u,v_i)+\sum_{d=1}^{2n-1}w_{{d}+1}(u)R^A_2(u,v_i)^{dd_1}_{d_1d}+w_{2n+1}(u)a^3_3(u,v_i)},\\
&&\beta_1(v_i)=T^{(d_1)}(v_i)\frac{w_{\bar{d}_1+1}(u)R^A_3(u,v_i,d_1)+w_{2n+1}(u)a^3_4(u,v_i,d_1)}
{w_{\bar{d}_1+1}(u)R^A_4(u,v_i,d_1)+w_{2n+1}(u)a^3_5(u,v_i,d_1)},\\
&&\sum_{l=1}^{2n+1}
w_l(u)\tilde{H}^{x_l}_{1,d_1}(u,v_i,v_j)-[\sum_{l=1}^{2n+1}
w_l(u)\tilde{H}^{x_l}_{2,d_1}(u,v_i,v_j)]\beta_1(v_i)\nonumber\\
&&-[\sum_{l=1}^{2n+1}
w_l(u)\tilde{H}^{x_l}_{3,d_1}(u,v_i,v_j)]\beta_1(v_j)+
[\sum_{l=1}^{2n+1}
w_l(u)\tilde{H}^{x_l}_{4,d_1}(u,v_i,v_j)]\beta_1(v_i)\beta_1(v_j)=0
\end{eqnarray}
with $x_1=A, x_{l+1}=\tilde{A}_{ll}, x_{2n+1}=\tilde{A}_2$,
$d_1=1,2,\cdots, 2n-1$.

\noindent From eqs.(\ref{taud}), (\ref{tupn1}) and (\ref{tau1}) ,
we can see
 that the diagonalization of $\tau(u)$ is reduced to finding
 the eigenvalue of $\tau_1(\tilde{u};\{\tilde{v}_m\})$ which is just
the transfer matrix of $A_{2(n-1)}^{(2)}$ vertex model with open
boundary conditions.
 Repeating the
procedure $j$ times, we can obtain the
$\Gamma_j({u}^{(j)};\{\tilde{v}_{m_{j-1}}^{(j-1)}\};\{{v}_{m_{j}}^{(j)}\})$
corresponding to the eigenvalue of open boundary
$A_{2(n-j)}^{(2)}$ vertex model,

\begin{eqnarray}
& & \hspace{-5mm}
\Gamma_j({u}^{(j)};\{\tilde{v}_{m_{j-1}}^{(j-1)}\};\{{v}_{m_{j}}^{(j)}\})\nonumber\\
&&=
w_1^{(j)}(u^{(j)})\omega_1^{(j)}(u^{(j)};\{\tilde{v}_{m_{j-1}}^{(j-1)}\}
)\Lambda_1^{m_j}(u^{(j)};\{{v}_{m_{j}}^{(j)}\})\nonumber\\
& &+
{w}^{(j)}(u^{(j)})\omega^{(j)}(u^{(j)};\{\tilde{v}_{m_{j-1}}^{(j-1)}\})
\Lambda_2^{m_j}(u^{(j)};\{{v}_{m_{j}}^{(j)}\})
\Gamma_{j+1}(u^{(j+1)};\{\tilde{v}_{m_{j}}^{(j)}\};\{{v}_{m_{j+1}}^{(j+1)}\})\nonumber\\
&
&+w_{2(n-j)+1}^{(j)}(u^{(j)})\omega_{2(n-j)+1}^{(j)}(u^{(j)};\{\tilde{v}_{m_{j-1}}^{(j-1)}\})
\Lambda_3^{m_j}(u^{(j)};\{{v}_{m_{j}}^{(j)}\}), \label{nesttu}
\end{eqnarray}
with ${u}^{(j)}={u}-{2j}\eta$,
 $\tilde{v}_{k}^{(j)}={v}_{k}^{(j)}-2\eta$,
 $\{{v}_{m_{j}}^{(j)}\}=\{v_1^{(j)}, \cdots,v_{m_j}^{(j)}\}$. $\{{v}_{m_{0}}^{(0)}\}= \{{v}_{m}\}$,
 $\{{v}_{m_{-1}}^{(-1)}\}=\{\tilde{v}_{m_{-1}}^{(-1)}\}= \{0\}$, $m_{-1}=N, m_0=m$.
 Replacing the $m, u, \{{v}_{m}\}, n$
 in the $\Lambda_l^{m}(u;\{{v}_{m}\})(l=1,2,3)$ by
 $m_j, u^{(j)}, \{{v}_{m_{j}}^{(j)}\}, n-j$ respectively, we have
  $\Lambda_l^{m_j}(u^{(j)};\{{v}_{m_{j}}^{(j)}\})$.
  The Bethe equations are
\begin{eqnarray}
&&\hspace{-8mm}\Gamma_{j+1}(\tilde{v}_{i}^{(j)};\{\tilde{v}_{m_{j}}^{(j)}\};\{{v}_{m_{j+1}}^{(j+1)}\})\nonumber\\
&&=-\rho^{-\frac{1}{2}
}_{n-j-1}(0)\frac{\omega_1^{(j)}(v^{(j)}_i;\{\tilde{v}_{m_{j-1}}^{(j-1)}\}
)\Lambda_1^{m_j-1}(v^{(j)}_i;\{\check{v}_{i}^{(j)}\})}
{\omega^{(j)}(v^{(j)}_i;\{\tilde{v}_{m_{j-1}}^{(j-1)}\}
)\Lambda_2^{m_j-1}(v^{(j)}_i;\{\check{v}_{i}^{(j)}\})}\beta_{j+1}(v_i^{(j)}).
\hspace{2mm}(i=1,\cdots,m_j)\label{nestBA}
\end{eqnarray}
The coefficients $w$'s, $\omega$'s, $\beta_j$ and the following
$\bar{\omega}$, $\xi$'s are expressed in Appendix B.  We can
rewrite the eigenvalue in detail, which is
\begin{eqnarray}
& & \hspace{-5mm} \Gamma({u})=\Gamma_0({u})=
{w}_1^{(0)}(u^{(0)})\bar{\omega}_1^{(0)}(u^{(0)} )
\xi_1^{(0)}(u^{(0)};\{\tilde{v}_{m_{-1}}^{(-1)}\}){\cal{A}}^{(m_0)}(u)\nonumber\\
&&+{w}_{2n+1}^{(0)}(u^{(0)})\bar{\omega}_{2n+1}^{(0)}(u^{(0)} )
\xi_3^{(0)}(u^{(0)};\{\tilde{v}_{m_{-1}}^{(-1)}\}){\cal{C}}^{(m_0)}(u)\nonumber\\
&&
+\sum_{j=0}^{n-2}\mu_j(u^{(j)})\nu_j(u^{(j)}){w}_1^{(j+1)}(u^{(j+1)})\bar{\omega}_1^{(j+1)}(u^{(j+1)}
)\xi_2^{(0)}(u^{(0)};\{\tilde{v}_{m_{-1}}^{(-1)}\}){\cal{B}}^{(m_j,m_{j+1})}(u)\nonumber\\
&&
+\sum_{j=0}^{n-2}\mu_j(u^{(j)})\nu_j(u^{(j)}){w}_{2(n-j-1)+1}^{(j+1)}(u^{(j+1)})\bar{\omega}_{2(n-j-1)+1}^{(j+1)}(u^{(j+1)}
)\xi_2^{(0)}(u^{(0)};\{\tilde{v}_{m_{-1}}^{(-1)}\})\bar{\cal{B}}^{(m_j,m_{j+1})}(u)\nonumber\\
&&+\mu_{n-1}(u^{(n-1)})\nu_{n-1}(u^{(n-1)})\xi_2^{(0)}(u^{(0)};\{\tilde{v}_{m_{-1}}^{(-1)}\}){\cal{{B}}}^{(m_{n-1})}(u),\label{nesttu1}
\end{eqnarray}
where
\begin{equation}
\mu_j(u^{(j)})=\prod_{i=0}^{j}\bar{\omega}^{(j)}(u^{(j)}),
\hspace{4mm}\nu_j(u^{(j)})=\prod_{i=0}^{j}{w}^{(j)}(u^{(j)}),
\end{equation}
\begin{eqnarray}
&&{\cal{A}}^{(m_0)}(u)=\prod_{k=1}^{m_{0}}\frac{\sinh(\frac{1}{2}({\tilde{u}^{(0)}+\tilde{v}^{(0)}_k})+2\eta)
\sinh(\frac{1}{2}({\tilde{u}^{(0)}-\tilde{v}^{(0)}_k})+2\eta)}
{\sinh(\frac{1}{2}({\tilde{u}^{(0)}+\tilde{v}^{(0)}_k}))
\sinh(\frac{1}{2}({\tilde{u}^{(0)}-\tilde{v}^{(0)}_k}))},\nonumber\\
&&{\cal{C}}^{(m_0)}(u)=\prod_{k=1}^{m_{0}}\frac{\cosh(\frac{1}{2}({\tilde{u}^{(0)}+\tilde{v}^{(0)}_k})-(2n+1)\eta)
\cosh(\frac{1}{2}({\tilde{u}^{(0)}-\tilde{v}^{(0)}_k})-(2n+1)\eta)}
{\cosh(\frac{1}{2}({\tilde{u}^{(0)}+\tilde{v}^{(0)}_k})-(2n-1)\eta)
\cosh(\frac{1}{2}({\tilde{u}^{(0)}-\tilde{v}^{(0)}_k})-(2n-1)\eta)},\nonumber\\
&&{\cal{B}}^{(m_j,m_{j+1})}(u)=\prod_{k=1}^{m_j}\frac{\sinh(\frac{1}{2}({\tilde{u}^{(j)}+\tilde{v}^{(j)}_k})-2\eta)
\sinh(\frac{1}{2}(\tilde{u}^{(j)}-\tilde{v}^{(j)}_k)-2\eta)}
{\sinh(\frac{1}{2}(\tilde{u}^{(j)}+\tilde{v}^{(j)}_k))\sinh(\frac{1}{2}(\tilde{u}^{(j)}-\tilde{v}^{(j)}_k))}\nonumber\\
&&\hspace{4mm}\times\prod_{l=1}^{m_{j+1}}\frac{\sinh(\frac{1}{2}({\tilde{u}^{(j+1)}+\tilde{v}^{(j+1)}_l})+2\eta)
\sinh(\frac{1}{2}({\tilde{u}^{(j+1)}-\tilde{v}^{(j+1)}_l})+2\eta)}
{\sinh(\frac{1}{2}({\tilde{u}^{(j+1)}+\tilde{v}^{(j+1)}_l}))\sinh(\frac{1}{2}({\tilde{u}^{(j+1)}-\tilde{v}^{(j+1)}_l}))},\nonumber\\
&&\bar{\cal{{B}}}^{(m_j,m_{j+1})}(u)=\prod_{k=1}^{m_j}\frac{\cosh(\frac{1}{2}({\tilde{u}^{(j)}+\tilde{v}^{(j)}_k})-(2n-2j-3)\eta)
\cosh(\frac{1}{2}(\tilde{u}^{(j)}-\tilde{v}^{(j)}_k)-(2n-2j-3)\eta)}
{\cosh(\frac{1}{2}(\tilde{u}^{(j)}+\tilde{v}^{(j)}_k)-(2n-2j-1)\eta)
\cosh(\frac{1}{2}(\tilde{u}^{(j)}-\tilde{v}^{(j)}_k)-(2n-2j-1)\eta)}\nonumber\\
&&\hspace{4mm}\times\prod_{l=1}^{m_{j+1}}\frac{\cosh(\frac{1}{2}({\tilde{u}^{(j+1)}+\tilde{v}^{(j+1)}_l})-(2n-2j-1)\eta)
\cosh(\frac{1}{2}({\tilde{u}^{(j+1)}-\tilde{v}^{(j+1)}_l})-(2n-2j-1)\eta)}
{\cosh(\frac{1}{2}({\tilde{u}^{(j+1)}+\tilde{v}^{(j+1)}_l})-(2n-2j-3)\eta)
\cosh(\frac{1}{2}({\tilde{u}^{(j+1)}-\tilde{v}^{(j+1)}_l})-(2n-2j-3)\eta)},\nonumber\\
&& \hspace{10cm}(j=0,1,2,\cdots,n-2)\nonumber\\
&&{\cal{{B}}}^{(m_{n-1})}(u)=\prod_{k=1}^{m_{n-1}}\frac{\sinh(\frac{1}{2}({\tilde{u}^{(n-1)}+\tilde{v}^{(n-1)}_k})-2\eta)
\sinh(\frac{1}{2}(\tilde{u}^{(n-1)}-\tilde{v}^{(n-1)}_k)-2\eta)}
{\sinh(\frac{1}{2}(\tilde{u}^{(n-1)}+\tilde{v}^{(n-1)}_k))\sinh(\frac{1}{2}(\tilde{u}^{(n-1)}-\tilde{v}^{(n-1)}_k))}\nonumber\\
&&\hspace{14mm}\times\frac{\cosh(\frac{1}{2}({\tilde{u}^{(n-1)}+\tilde{v}^{(n-1)}_k})+\eta)
\cosh(\frac{1}{2}(\tilde{u}^{(n-1)}-\tilde{v}^{(n-1)}_k)+\eta)}
{\cosh(\frac{1}{2}(\tilde{u}^{(n-1)}+\tilde{v}^{(n-1)}_k)-\eta)
\cosh(\frac{1}{2}(\tilde{u}^{(n-1)}-\tilde{v}^{(n-1)}_k)-\eta)}.
\end{eqnarray}
The explicit expression of Bethe equations eq.(\ref{nestBA}) is
\begin{eqnarray}
\displaystyle&&\prod_{k=1}^{m_{j-1}}\frac{\sinh(\frac{1}{2}({\tilde{v}_i^{(j)}+\tilde{v}_k^{(j-1)}})-\eta)
\sinh(\frac{1}{2}({\tilde{v}_i^{(j)}-\tilde{v}_k^{(j-1)}})-\eta)}
{\sinh(\frac{1}{2}({\tilde{v}_i^{(j)}+\tilde{v}_k^{(j-1)}})+\eta)
\sinh(\frac{1}{2}({\tilde{v}_i^{(j)}-\tilde{v}_k^{(j-1)}})+\eta)}\nonumber\\
&&\times\prod_{l=1}^{m_{j+1}}\frac{\sinh(\frac{1}{2}({\tilde{v}_i^{(j)}+\tilde{v}_l^{(j+1)}})-\eta)
\sinh(\frac{1}{2}({\tilde{v}_i^{(j)}-\tilde{v}_l^{(j+1)}})-\eta)}
{\sinh(\frac{1}{2}({\tilde{v}_i^{(j)}+\tilde{v}_l^{(j+1)}})+\eta)
\sinh(\frac{1}{2}({\tilde{v}_i^{(j)}-\tilde{v}_l^{(j+1)}})+\eta)}\nonumber\\
&&\times\prod_{s=1\ne i
}^{m_{j}}\frac{\sinh(\frac{1}{2}({\tilde{v}_i^{(j)}+\tilde{v}_s^{(j)}})+2\eta)
\sinh(\frac{1}{2}({\tilde{v}_i^{(j)}+\tilde{v}_s^{(j)}})+2\eta)}
{\sinh(\frac{1}{2}({\tilde{v}_i^{(j)}+\tilde{v}_s^{(j)}})-2\eta)
\sinh(\frac{1}{2}({\tilde{v}_i^{(j)}+\tilde{v}_s^{(j)}})-2\eta))}\nonumber\\
&&=W^{(j)}(\tilde{v}_i^{(j)})\Omega^{(j)}(\tilde{v}_i^{(j)}),
\hspace{2mm}(i=1,\cdots,m_j; j\ne n-1)\label{nestBA1}
\end{eqnarray}
\begin{eqnarray}
&&\prod_{k=1}^{m_{n-2}}\frac{\sinh(\frac{1}{2}({\tilde{v}_i^{(n-1)}+\tilde{v}_k^{(n-2)}})-\eta)
\sinh(\frac{1}{2}({\tilde{v}_i^{(n-1)}-\tilde{v}_k^{(n-2)}})-\eta)}
{\sinh(\frac{1}{2}({\tilde{v}_i^{(n-1)}+\tilde{v}_k^{(n-2)}})+\eta)
\sinh(\frac{1}{2}({\tilde{v}_i^{(n-1)}-\tilde{v}_k^{(n-2)}})+\eta)}\nonumber\\
&&\prod_{l=1}^{m_{n-1}}\frac{\cosh(\frac{1}{2}({\tilde{v}_i^{(n-1)}+\tilde{v}_l^{(n-1)}})-\eta)
\cosh(\frac{1}{2}({\tilde{v}_i^{(n-1)}-\tilde{v}_l^{(n-1)}})-\eta)}
{\cosh(\frac{1}{2}({\tilde{v}_i^{(n-1)}+\tilde{v}_l^{(n-1)}})+\eta)
\cosh(\frac{1}{2}({\tilde{v}_i^{(n-1)}-\tilde{v}_l^{(n-1)}})+\eta)}\nonumber\\
&&\hspace{3mm}\times\frac{\sinh(\frac{1}{2}({\tilde{v}_i^{(n-1)}+\tilde{v}_l^{(n-1)}})+2\eta)
\sinh(\frac{1}{2}({\tilde{v}_i^{(n-1)}+\tilde{v}_l^{(n-1)}})+2\eta)}
{\sinh(\frac{1}{2}({\tilde{v}_i^{(n-1)}+\tilde{v}_l^{(n-1)}})-2\eta)
\sinh(\frac{1}{2}({\tilde{v}_i^{(n-1)}+\tilde{v}_l^{(n-1)}})-2\eta))}\nonumber\\
&&=W^{(n-1)}(\tilde{v}_i^{(n-1)})
\Omega^{(n-1)}(\tilde{v}_i^{(n-1)})
\hspace{2mm}(i=1,\cdots,m_{n-1})\label{nestBA2}
\end{eqnarray}
with
\begin{eqnarray}
&&W^{(j)}(\tilde{v}_i^{(j)})=\displaystyle\left\{\begin{array}{ll}\displaystyle-\frac{{w}_1^{(j+1)}(\tilde{v}_i^{(j)})a_{n-j-1}(2\tilde{v}_i^{(j)}))
}{\beta_{j+1}(v_i^{(j)})},& j\ne n-1\\[2mm]
\displaystyle-\frac{1}{\beta_{n}(v_i^{(n-1)})},& j=n-1\end{array}\right.\\
&&\Omega^{(j)}(\tilde{v}_i^{(j)})=\left\{\begin{array}{ll}
\displaystyle\frac{\bar{\omega}^{(j)}(v_i^{(j)})\bar{\omega}_1^{(j+1)}(\tilde{v}_i^{(j)})}
{\bar{\omega}_1^{(j)}({v}_i^{(j)})},& j\ne n-1\\[2mm]
\displaystyle\frac{\bar{\omega}^{(j)}(v_i^{(n-1)})}
{\bar{\omega}_1^{(n-1)}({v}_i^{(n-1)})}.&j=n-1\end{array}\right.
\end{eqnarray}
 Up to now, we have gotten the whole eigenvalues and the
Bethe equations of transfer matrix   for the $A_{2n}^{(2)}$ vertex
model with open boundary condition.

\section{Conclusions}

In the framework  of algebraic Bethe ansatz, we solve the
$A_{2n}^{(2)}$ vertex model with general diagonal reflecting
matrices. When the model is $U_q(B_n)$ quantum invariant, we find
that our conclusion agrees with that obtained by analytic Bethe
ansatz method \cite{a2n2}. It seems that other models, such as
$A^{(2)}_{2n-1}$, $B^{(1)}_n$ and  $C^{(1)}_n$ vertex models can
also be treated in this way. Additionally, we notice that
algebraic Bethe ansatz  has been generalized to the spin chain
with non-diagonal reflecting matrices\cite{cao}-\cite{yangwl}.  It
is interesting to apply the method to other higher rank algebras.

\section*{ Acknowledgements}

This work is supported by  NNSFC under Grant No.10175050 and
No.90403019, and Program for  NCET
under Grant No. NCET-04-0929.

\setcounter{section}{0}
\renewcommand{\thesection}{\Alph{section}}

\section{Some detail derivations }
\setcounter{equation}{0}
\renewcommand{\theequation}{A.\arabic{equation}}

Acting the diagonal operators
$x(u)=A(u),\tilde{A}_{aa}(u),\tilde{A}_2(u)$ on the m-particle
state and having carried out a very involved analysis similar to
that in Ref.\cite{ik}, we can obtain the following expression
\begin{eqnarray}
& &
\hspace{-10mm}{x}(u)|\Upsilon_m(v_1,\cdots,v_m)\rangle=|\Psi_{x}(u,\{v_m\})\rangle\nonumber\\
 & &+\sum_{i=1}^m h_1^x(u,v_i,d_1)|\Psi_{m-1}^{(1)}(u,v_i;\{v_m\})_{d_1d_1}\rangle\nonumber\\
& & +\sum_{i=1}^m  h_2^x(u,v_i,d)|\Psi_{m-1}^{(2)}(u,v_i;\{v_m\})_{dd}\rangle\nonumber\\
& &+\sum_{i=1}^m h_3^x(u,v_i,\bar{\alpha}_x)|\Psi_{m-1}^{(3)}(u,v_i;\{v_m\})_{\alpha_x\alpha_x}\rangle\nonumber\\
& &+\sum_{i=1}^m h_4^x(u,v_i,\bar{\alpha}_x)|\Psi_{m-1}^{(4)}(u,v_i;\{v_m\})_{\alpha_x\alpha_x}\rangle\nonumber\\
& & +\sum^{m-1}_{i=1}\sum_{j=i+1}^m \delta_{\bar{d}_1e_2}
H^{{x}}_{1,d_1}(u,v_i,v_j)|\Psi_{m-2}^{(5)}(u,v_i,v_j;\{v_m\})_{d_1e_2}\rangle\nonumber\\
& & +\sum^{m-1}_{i=1}\sum_{j=i+1}^m
H^{{x}}_{2,d_1}(u,v_i,v_j)^{fe}_{c_2d_1}|\Psi_{m-2}^{(6)}(u,v_i,v_j;\{v_m\})_{d_1c_2}^{ef}\rangle\nonumber\\
& & +\sum^{m-1}_{i=1}\sum_{j=i+1}^m H^{{x}}_{3,d_1}(u,v_i,v_j)
|\Psi_{m-2}^{(7)}(u,v_i,v_j;\{v_m\})_{d_1d_1}\rangle\nonumber\\
& & +\sum^{m-1}_{i=1}\sum_{j=i+1}^m
H^{{x}}_{4,d_1}(u,v_i,v_j)^{de}_{fd_1}|\Psi_{m-2}^{(8)}(u,v_i,v_j;\{v_m\})_{d_1f}^{ed}\rangle,
\label{d1pns}
\end{eqnarray}
where when $x=A,\tilde{A}_{aa},\tilde{A}_2$,
\begin{eqnarray}
&&h_1^x(u,v_i,d_1)=a^1_2(u,v_i),
R^A_1(u,v_i)^{d_1a}_{ad_1},a^3_2(u,v_i),\nonumber\\
&&h_2^x(u,v_i,d)=a^1_3(u,v_i),
R^A_2(u,v_i)^{da}_{ad},a^3_3(u,v_i),\nonumber\\
&&h_3^x(u,v_i,\bar{\alpha}_x)=0,
R^A_3(u,v_i,\bar{a}),a^3_4(u,v_i,\bar{d}),\nonumber\\
&&h_4^x(u,v_i,\bar{\alpha}_x)=0,
R^A_4(u,v_i,\bar{a}),a^3_5(u,v_i,\bar{d}),\nonumber\\
&&\alpha_x=0, a, d, \nonumber
\end{eqnarray}
\begin{eqnarray}
|\Psi_{x}(u,\{v_m\})\rangle=\left\{\begin{array}{l}\omega_1(u)\Lambda_1^m(u;v_1,\cdots,v_m)
|\Upsilon_m(v_1,\cdots,v_m)\rangle,\\[2mm]
\Phi_m^{d_1\cdots
d_m}(v_1,\cdots,v_m)[\tilde{T}^{m}(u;\{{v}_m\})^{d_1\cdots
d_m}_{b_1\cdots b_m}]_{aa}F^{b_1\cdots b_m}|0\rangle,\\[2mm]
\omega_{2n+1}(u)\Lambda_3^m(u;v_1,\cdots,v_m)
|\Upsilon_m(v_1,\cdots,v_m)\rangle, \end{array}\right.
\end{eqnarray}
respectively, and we denote
\begin{eqnarray}
& & \hspace{-6mm}|\Psi_{m-1}^{(1)}(u,v_i;\{v_m\})_{fd_1}\rangle=
B_f(u)|\Phi_{m-1}^{d_2\cdots
d_m}(v_1,\cdots,\check{v}_i,\cdots,v_m) \nonumber \\
& & \hspace{8mm}\times S^{d_1\cdots d_m}_{b_1\cdots
b_m}(v_i;\{\check{v}_i\})\Lambda_1^{m-1}(v_i; \{\check{v}_i\})
\omega_1(v_i)F^{b_1\cdots b_m}|0\rangle,\\
& &
\hspace{-6mm}|\Psi_{m-1}^{(2)}(u,v_i;\{v_m\})_{fd}\rangle=B_f(u)\Phi_{m-1}^{e_2\cdots
e_m}(v_1,\cdots,\check{v}_i,\cdots,v_m)\nonumber\\
& & \hspace{10mm}
\times[\tilde{T}^{m-1}(v_i;\{\check{v}_i\})^{e_2\cdots
e_m}_{d_2'\cdots d_m'}]_{dd_1'}  S^{d_1'\cdots d_m'}_{b_1\cdots
b_m}(v_i;\{\check{v}_i\})F^{b_1\cdots b_m}|0\rangle,\\
& & \hspace{-6mm}|\Psi_{m-1}^{(3)}(u,v_i;\{v_m\})_{ab}\rangle=
E_{{a}}(u)\Phi_{m-1}^{d_2\cdots
d_m}(v_1,\cdots,\check{v}_i,\cdots,v_m) \nonumber\\
& & \hspace{10mm} \times S^{d_1\cdots d_m}_{b_1\cdots
b_m}(v_i;\{\check{v}_i\})\Lambda_1^{m-1}(v_i; \{\check{v}_i\})
\omega_1(v_i) \delta_{\bar{b}d_1}F^{b_1\cdots b_m}|0\rangle,\\
& & \hspace{-6mm}
|\Psi_{m-1}^{(4)}(u,v_i;\{v_m\})_{ab}\rangle=E_a(u)\Phi_{m-1}^{e_2\cdots
e_m}(v_1,\cdots,\check{v}_i,\cdots,v_m)\nonumber\\
& & \hspace{10mm}\times
[\tilde{T}^{m-1}(v_i;\{\check{v}_i\})^{e_2\cdots e_m}_{d_2'\cdots
d_m'}]_{\bar{b}d_1'} S^{d_1'\cdots d_m'}_{b_1\cdots
b_m}(v_i;\{\check{v}_i\}) F^{b_1\cdots b_m}|0\rangle,\\
& &\hspace{-6mm}
|\Psi_{m-2}^{(5)}(u,v_i,v_j;\{v_m\})_{d_1e_2}\rangle =
F(u)\Phi_{m-2}^{e_3\cdots e_m}(v_1,\cdots,\check{v}_i,\cdots,\check{v}_j,\cdots,v_m)\nonumber \\
& & \hspace{4mm}\times S^{e_2\cdots e_m}_{d_2\cdots
d_m}(v_j;\{\check{v}_i,\check{v}_j\})S^{d_1\cdots d_m}_{b_1\cdots
b_m}(v_i;\{\check{v}_i\})\Lambda_1^{m-2}(v_i;
\{\check{v}_i,\check{v}_j\})\nonumber \\
& & \hspace{4mm}\times \Lambda_1^{m-2}(v_j;
\{\check{v}_i,\check{v}_j\})A(v_i)A(v_j)F^{b_1\cdots
b_m}|0\rangle,\\
& &\hspace{-6mm}
|\Psi_{m-2}^{(6)}(u,v_i,v_j;\{v_m\})_{d_1c_2}^{ef}\rangle =
F(u)\Phi_{m-2}^{e_3\cdots e_m}(v_1,\cdots,\check{v}_i,\cdots,\check{v}_j,\cdots,v_m)\nonumber \\
& & \hspace{4mm}\times
[\tilde{T}^{m-2}(v_i;\{\check{v}_i,\check{v}_j\})^{e_3\cdots
e_m}_{c_3\cdots c_m}]_{\bar{f}e} S^{c_2\cdots c_m}_{d_2\cdots
d_m}(v_j;\{\check{v}_i,\check{v}_j\})S^{d_1\cdots d_m}_{b_1\cdots
b_m}(v_i;\{\check{v}_i\})\nonumber \\
& & \hspace{4mm}\times \Lambda_1^{m-2}(v_j;
\{\check{v}_i,\check{v}_j\})A(v_j)F^{b_1\cdots b_m}|0\rangle,\\
& &\hspace{-6mm}
|\Psi_{m-2}^{(7)}(u,v_i,v_j;\{v_m\})_{fd_1}\rangle =
F(u)\Phi_{m-2}^{e_3\cdots e_m}(v_1,\cdots,\check{v}_i,\cdots,\check{v}_j,\cdots,v_m)\nonumber \\
& & \hspace{4mm}\times
[\tilde{T}^{m-2}(v_j;\{\check{v}_i,\check{v}_j\})^{e_3\cdots
e_m}_{c_3\cdots c_m}]_{\bar{f}c_2} S^{c_2\cdots c_m}_{d_2\cdots
d_m}(v_j;\{\check{v}_i,\check{v}_j\})S^{d_1\cdots d_m}_{b_1\cdots
b_m}(v_i;\{\check{v}_i\})\nonumber \\
& & \hspace{4mm}\times \Lambda_1^{m-2}(v_i;
\{\check{v}_i,\check{v}_j\})A(v_i)F^{b_1\cdots b_m}|0\rangle,\\
& &\hspace{-6mm}
|\Psi_{m-2}^{(8)}(u,v_i,v_j;\{v_m\})_{d_1f}^{ed}\rangle =
F(u)\Phi_{m-2}^{e_3\cdots e_m}(v_1,\cdots,\check{v}_i,\cdots,\check{v}_j,\cdots,v_m)\nonumber \\
& & \hspace{4mm}\times
[\tilde{T}^{m-2}(v_i;\{\check{v}_i,\check{v}_j\})^{e_3\cdots
e_m}_{a_3\cdots a_m}]_{\bar{d}e}
[\tilde{T}^{m-2}(v_j;\{\check{v}_i,\check{v}_j\})^{a_3\cdots
a_m}_{c_3\cdots c_m}]_{fc_2}\nonumber \\
& & \hspace{4mm}\times S^{c_2\cdots c_m}_{d_2\cdots
d_m}(v_j;\{\check{v}_i,\check{v}_j\})S^{d_1\cdots d_m}_{b_1\cdots
b_m}(v_i;\{\check{v}_i\}) F^{b_1\cdots b_m}|0\rangle.
\end{eqnarray}
The explicit expressions of $H^{x}_{l,d_1}(u,v_i,v_j), l=1,2,3,4$
are listed as below
\begin{eqnarray}
 & & \hspace{-6mm} H^{A}_{1,d_1}(u,v_i,v_j)= a^1_4(u,v_i,\bar{d}_1)(c^1_5(v_i,v_j)+c^1_7(v_i,v_j))
  \nonumber\\
& &+
a^1_5(u,v_i)(c^2_4(v_i,v_j,\bar{d}_1)+c^2_6(v_i,v_j,\bar{d}_1))+b^1_2(u,v_i)g_{1}(v_i,v_j,d_1)\nonumber\\
& & +a^1_1(u,v_i)a^1_2(u,v_j)g_1(u,v_i,d)
\hat{r}(v_i-u)^{\bar{d}d}_{d_1\bar{d}_1},\\[2mm]
 & & \hspace{-6mm} H^{A}_{2,d_1}(u,v_i,v_j)^{fe}_{c_2d_1}=
 a^1_4(u,v_i,\bar{d}_1)(c^1_6(v_i,v_j)+c^1_9(v_i,v_j))\delta_{d_1f}
  \nonumber\\
& &+
a^1_5(u,v_i)(R^C_3(v_i,v_j)^{fe}_{c_2d_1}+R^C_4(v_i,v_j)^{fe}_{c_2d_1})\nonumber\\
& &+b^1_3(u,v_i)g_1(v_i,v_j,d_1)\delta_{d_1\bar{c}_2}+
a^1_1(u,v_i)a^1_2(u,v_j)g_2(u,v_i,f) \hat{r}(v_i-u)^{fe}_{c_2d_1},\\[2mm]
 & & \hspace{-6mm} H^{A}_{3,d_1}(u,v_i,v_j)= a^1_4(u,v_i,\bar{d}_1)c^1_8(v_i,v_j)
  + a^1_5(u,v_i)c^2_7(v_i,v_j,\bar{d}_1)\nonumber\\
&
&+b^1_2(u,v_i)g_{2}(v_i,v_j,d_1)+a^1_1(u,v_i)a^1_3(u,v_j)g_1(u,v_i,d)
\hat{r}(v_i-u)^{d\bar{d}}_{\bar{d}_1d_1},\\[2mm]
 & & \hspace{-6mm} H^{A}_{4,d_1}(u,v_i,v_j)^{de}_{fd_1}=
 a^1_4(u,v_i,\bar{d}_1)c^1_{10}(v_i,v_j)\delta_{d_1d}+
a^1_5(u,v_i)R^C_5(v_i,v_j)^{de}_{fd_1}\nonumber\\
& &+b^1_3(u,v_i)g_2(v_i,v_j,d_1)\delta_{\bar{d}_1f}+
a^1_1(u,v_i)a^1_3(u,v_j)g_2(u,v_i,d) \hat{r}(v_i-u)^{de}_{fd_1},\\[2mm]
 & & \hspace{-6mm} H^{\tilde{A}_{aa}}_{1,d_1}(u,v_i,v_j)= R^A_5(u,v_i)^{ad_1}_{d_1a}(c^1_5(v_i,v_j)+c^1_7(v_i,v_j))
  +b^2_2(u,v_i)g_{1}(v_i,v_j,d_1)\nonumber\\
& &+
R^A_6(u,v_i)(c^2_4(v_i,v_j,\bar{d}_1)+c^2_6(v_i,v_j,\bar{d}_1))
+\tilde{r}(u+v_i)^{ad}_{da}\bar{r}(u-v_i)^{ad_1}_{d_1a}R^A_3(u,v_j,\bar{d}_1)e^1_1(v_i,u,d)\nonumber\\
& &
+\tilde{r}(u+v_i)^{ae}_{dg}\bar{r}(u-v_i)^{gf}_{d_1a}g_1(u,v_i,h)R^A_{1}(u,v_j)^{d\bar{e}}_{\bar{d}_1f}
\hat{r}(v_i-u)^{h\bar{h}}_{\bar{e}e},\\[2mm]
 & & \hspace{-6mm} H^{\tilde{A}_{aa}}_{2,d_1}(u,v_i,v_j)^{fe}_{c_2d_1}=
 R^A_5(u,v_i)^{ad_1}_{d_1a}(c^1_6(v_i,v_j)+c^1_9(v_i,v_j))\delta_{d_1f}
  \nonumber\\
& &+
R^A_6(u,v_i)(R^C_3(v_i,v_j)^{fe}_{c_2d_1}+R^C_4(v_i,v_j)^{fe}_{c_2d_1})
+R^F_1(u,v_i)^{fe}_{a\bar{a}}g_1(v_i,v_j,d_1)\delta_{d_1\bar{c}_2}\nonumber\\
& &+
\tilde{r}(u+v_i)^{ac}_{dg}\bar{r}(u-v_i)^{gh}_{d_1a}g_2(u,v_i,f)R^A_{1}(u,v_j)^{db}_{c_2h}
\hat{r}(v_i-u)^{fe}_{bc}\nonumber\\
& &
+\tilde{r}(u+v_i)^{ah}_{dg}\bar{r}(u-v_i)^{g\bar{c}_2}_{d_1a}R^A_{3}(u,v_j,\bar{c}_2)
R^{be}_3(v_i,u)^{fe}_{\bar{d}h},\\[2mm]
 & & \hspace{-6mm} H^{\tilde{A}_{aa}}_{3,d_1}(u,v_i,v_j)= R^A_5(u,v_i)^{ad_1}_{d_1a}c^1_8(v_i,v_j)
  + R^A_6(u,v_i)c^2_7(v_i,v_j,\bar{d}_1)\nonumber\\
&
&+b^2_2(u,v_i)g_{2}(v_i,v_j,d_1)+\tilde{r}(u+v_i)^{ad}_{da}\bar{r}(u-v_i)^{ad_1}_{d_1a}R^A_4(u,v_j,\bar{d}_1)e^1_1(v_i,u,d)\nonumber\\
& &
+\tilde{r}(u+v_i)^{ae}_{dg}\bar{r}(u-v_i)^{gf}_{d_1a}g_1(u,v_i,h)R^A_{2}(u,v_j)^{d\bar{e}}_{\bar{d}_1f}
\hat{r}(v_i-u)^{h\bar{h}}_{\bar{e}e},\\[2mm]
 & & \hspace{-6mm} H^{\tilde{A}_{aa}}_{4,d_1}(u,v_i,v_j)^{de}_{fd_1}=
 R^A_5(u,v_i)^{ad_1}_{d_1a}c^1_{10}(v_i,v_j)\delta_{d_1d}+
R^A_6(u,v_i)R^C_5(v_i,v_j)^{de}_{fd_1}\nonumber\\
&
&+R^F_1(u,v_i)^{de}_{a\bar{a}}g_2(v_i,v_j,d_1)\delta_{d_1\bar{f}}
+\tilde{r}(u+v_i)^{ac}_{hg}\bar{r}(u-v_i)^{g\bar{f}}_{d_1a}R^A_{4}(u,v_j,f)
R^{be}_3(v_i,u)^{de}_{\bar{h}c}\nonumber\\
& & +
\tilde{r}(u+v_i)^{ac}_{bg}\bar{r}(u-v_i)^{gh}_{d_1a}g_2(u,v_i,d)R^A_{2}(u,v_j)^{bk}_{fh}
\hat{r}(v_i-u)^{de}_{kc},\\[2mm]
 & & \hspace{-6mm} H^{\tilde{A}_{2}}_{1,d_1}(u,v_i,v_j)= a^3_6(u,v_i,\bar{d}_1)(c^1_5(v_i,v_j)+c^1_7(v_i,v_j))
  +b^3_2(u,v_i)g_{1}(v_i,v_j,d_1)\nonumber\\
& &+
a^3_7(u,v_i)(c^2_4(v_i,v_j,\bar{d}_1)+c^2_6(v_i,v_j,\bar{d}_1))
+a^3_1(u,v_i)a^3_4(u,v_j,\bar{d}_1)e^1_1(v_i,u,d_1)\nonumber\\
& & +a^3_1(u,v_i)a^3_2(u,v_j)g_1(u,v_i,d)
\hat{r}(v_i-u)^{d\bar{d}}_{\bar{d}_1d_1},\\[2mm]
 & & \hspace{-6mm} H^{\tilde{A}_{2}}_{2,d_1}(u,v_i,v_j)^{fe}_{c_2d_1}=
 a^3_6(u,v_i,\bar{d}_1)(c^1_6(v_i,v_j)+c^1_9(v_i,v_j))\delta_{d_1f}
  \nonumber\\
& &+
a^3_7(u,v_i)(R^C_3(v_i,v_j)^{fe}_{c_2d_1}+R^C_4(v_i,v_j)^{fe}_{c_2d_1})\nonumber\\
&
&+b^3_3(u,v_i)g_1(v_i,v_j,d_1)\delta_{d_1\bar{c}_2}+a^3_1(u,v_i)a^3_4(u,v_j,c_2)
R^{be}_3(v_i,u)^{fe}_{c_2d_1}\nonumber\\
& & + a^3_1(u,v_i)a^3_2(u,v_j)g_2(u,v_i,f)
\hat{r}(v_i-u)^{fe}_{c_2d_1},\\[2mm]
 & & \hspace{-6mm} H^{\tilde{A}_{2}}_{3,d_1}(u,v_i,v_j)= a^3_6(u,v_i,\bar{d}_1)c^1_8(v_i,v_j)
  + a^3_7(u,v_i)c^2_7(v_i,v_j,\bar{d}_1)\nonumber\\
&
&+b^3_2(u,v_i)g_{2}(v_i,v_j,d_1)+a^3_1(u,v_i)a^3_5(u,v_j,d_1)e^1_1(v_i,u,d_1)\nonumber\\
& & +a^3_1(u,v_i)a^3_3(u,v_j)g_1(u,v_i,d)
\hat{r}(v_i-u)^{d\bar{d}}_{\bar{d}_1d_1},\\[2mm]
 & & \hspace{-6mm} H^{\tilde{A}_{2}}_{4,d_1}(u,v_i,v_j)^{de}_{fd_1}=
 a^3_6(u,v_i,\bar{d}_1)c^1_{10}(v_i,v_j)\delta_{d_1d}+
a^3_7(u,v_i)R^C_5(v_i,v_j)^{de}_{fd_1}\nonumber\\
& &+b^3_3(u,v_i)g_2(v_i,v_j,d_1)\delta_{d_1\bar{f}}
+a^3_1(u,v_i)a^3_5(u,v_j,\bar{f}) R^{be}_3(v_i,u)^{de}_{fd_1}\nonumber\\
& &+ a^3_1(u,v_i)a^3_3(u,v_j)g_2(u,v_i,d)
\hat{r}(v_i-u)^{de}_{fd_1}.
\end{eqnarray}
 All the repeated indices sum over 1 to $2n-1$ except for $a,
d_1$ and we have checked that
\begin{eqnarray}
\frac{H^{x}_{2,b}(u,v_i,v_j)^{{d}_1a}_{cb}}{R^{(n-1)}(\tilde{v}_i-\tilde{v}_j)^{{d}_1a}_{cb}}=
\frac{H^{x}_{2,{d}_1}(u,v_i,v_j)^{{d}_1{d}_1}_{{d}_1{d}_1}}{R^{(n-1)}(\tilde{v}_i-\tilde{v}_j)^{{d}_1{d}_1}_{{d}_1{d}_1}},
\label{haa2}
\end{eqnarray}
\begin{eqnarray}
\frac{H^{x}_{4,b}(u,v_i,v_j)^{{d}_1a}_{cb}}{R^{(n-1)}(\tilde{v}_i+\tilde{v}_j)^{{d}_1a}_{cb}}=
\frac{H^{x}_{4,{d}_1}(u,v_i,v_j)^{{d}_1{d}_1}_{{d}_1{d}_1}}{R^{(n-1)}(\tilde{v}_i+\tilde{v}_j)^{{d}_1{d}_1}_{{d}_1{d}_1}}.
\label{haa4}
\end{eqnarray}

 \noindent We conclude that  Eq.(\ref{d1pns}) can be verified
directly by using mathematical induction, although it is a rather
hard work. Similar to assumption of algebraic Bethe ansatz, we
might assume that ``quasi'' m-particle states such as
$B\Phi_{m-1}|0\rangle$, $E\Phi_{m-1}|0\rangle$,
$BB\Phi_{m-2}|0\rangle$, $BE\Phi_{m-2}|0\rangle$,
$EB\Phi_{m-2}|0\rangle$, $F\Phi_{m-2}|0\rangle$,
$FB\Phi_{m-3}|0\rangle$ etc are linearly independent. Here
 all the indices are omitted and all the spectrum parameters in the
``quasi'' n-particle state keep the order
$\{v_{i_1},v_{i_2},\cdots,v_{i_k}\}$ with\  $i_1 <i_2 < \cdots
<i_k$. For example, $B_{1}(v_1) B_{1}(v_2) \Phi_{m-2}^{b_3\cdots
b_m}$ $(v_3,\cdots,v_m) F^{11b_3\cdots b_m}|0\rangle$ and
$B_{1}(v_1)B_{2}(v_2)\Phi_{m-2}^{b_3\cdots
b_m}(v_3,\cdots,v_m)F^{12b_3\cdots b_m}|0\rangle$  are thought to
be linearly independent. Then, by using the assumption, the
property of Eq.(\ref{npsp}) and some necessary relations, we can
prove the  conclusions Eq.(\ref{d1pns}) as done in Ref.\cite{ik}.

In order to obtain the eigenvalue and the corresponding Behe
equations, we need to carry on the following procedure. Denote
\begin{eqnarray}
&&[\tilde{T}^{m}(u;\{{v}_m\})^{d_1\cdots d_m}_{b_1\cdots
b_m}]_{ab} F^{b_1\cdots
b_m}|0\rangle=\omega(u)\Lambda_2^m(u;\{{v}_m\})[{T}^{m}(u;\{{v}_m\})^{d_1\cdots
d_m}_{b_1\cdots b_m}]_{ab} F^{b_1\cdots b_m}|0\rangle\nonumber\\
&&\Lambda_2^{m}(u; \{{v}_m\})=\prod
_{i=1}^m \rho_{n-1}(u-v_i)\tilde{\rho}(u,v_i),\\
&&\rho_{n-1}(u)=a_{n-1}(u)a_{n-1}(-u),\hspace{4mm}
 \tilde{\rho}(u,v)=\frac{1}{a_n(u+v)e_n(u-v)},\\
& &[T^{m}(u;\{{v}_m\})^{d_1\cdots d_m}_{c_1\cdots
c_m}]_{ab}\nonumber \\
&&={L}(\tilde{u},\tilde{v}_1)^{ad_1}_{h_1g_1}
   {L}(\tilde{u},\tilde{v}_2)^{h_1d_2}_{h_2g_2}\cdots
   {L}(\tilde{u},\tilde{v}_m)^{h_{m-1}d_m}_{h_mg_m}k^-(u)_{h_m}\nonumber\\
& & \hspace{2mm}\times
{L^{-1}}(-\tilde{u},\tilde{v}_m)^{h_mg_m}_{f_{m-1}c_m}
{L^{-1}}(-\tilde{u},\tilde{v}_{m-1})^{f_{m-1}g_{m-1}}_{f_{m-2}c_{m-1}}
\cdots {L^{-1}}(-\tilde{u},\tilde{v}_1)^{f_1g_1}_{bc_1}.
\label{trans2}
\end{eqnarray}
Before deducing the Eq.(\ref{nd1pns}), we present the following
four relations (the proofs are omitted here)
\begin{eqnarray}
& & \hspace{-6mm}S^{d_1\cdots d_m}_{c_1\cdots
c_m}(v_i;\{\check{v}_i\})\tau_1(\tilde{v}_i;\{\tilde{v}_m\})^{c_1\cdots
c_m}_{b_1\cdots b_m}=\nonumber \\
& & ({\rho^{\frac{1}{2}}_{n-1}(0)})^{-1}T^{(d_1)}(v_i)
[T^{m-1}(v_i;\{\check{v}_i\})^{d_2\cdots d_m}_{c_2'\cdots
c_m'}]_{d_1c_1'}S^{c_1'\cdots c_m'}_{b_1\cdots
b_m}(v_i;\{\check{v}_i\}), \label{ir1}\\[2mm]
& &\hspace{-6mm}
R^{(n-1)}(\tilde{v}_i-\tilde{v}_j)^{\bar{a}_1h_1'}_{c_2'd_1'}[{T}^{m-2}(v_i;\{\check{v}_i,\check{v}_j\})^{e_3\cdots
e_m}_{c_3'\cdots c_m'}]_{a_1h_1'} S^{c_2'\cdots c_m'}_{d_2'\cdots
d_m'}(v_j;\{\check{v}_i,\check{v}_j\})S^{d_1'\cdots
d_m'}_{b_1\cdots b_m}(v_i;\{\check{v}_i\})\nonumber \\
&
&\hspace{3mm}=\frac{\rho_{n-1}(\tilde{v}_i-\tilde{v}_j){\rho^{\frac{1}{2}}_{n-1}(0)}}
{\rho_{n-1}(\tilde{v}_i+\tilde{v}_j)}
\frac{R^{(n-1)}(-\tilde{v}_i-\tilde{v}_j)^{\bar{a}_1a_1}_{\bar{d}d}}{T^{(d)}(v_i)}\nonumber \\
& &\hspace{6mm}\times S^{\bar{d}e_3\cdots e_m}_{h_2\cdots
h_m}(v_j;\{\check{v}_i,\check{v}_j\})S^{{d}h_2\cdots
h_m}_{d_1\cdots
d_m}(v_i;\{\check{v}_i\})]\tau_1(\tilde{v}_i;\{\tilde{v}_m\})^{d_1\cdots
d_m}_{b_1\cdots b_m},\label{fu22}\\
& &\hspace{-8mm}
[{T}^{m-2}(v_j;\{\check{v}_i,\check{v}_j\})^{e_3\cdots
e_m}_{c_3'\cdots c_m'}]_{\bar{c}_1c_2'}S^{c_2'\cdots
c_m'}_{d_2'\cdots
d_m'}(v_j;\{\check{v}_i,\check{v}_j\})S^{{c}_1d_2'\cdots
d_m'}_{b_1\cdots
b_m}(v_i;\{\check{v}_i\})\nonumber \\
& & \hspace{-2mm} =\frac{{\rho^{\frac{1}{2}}_{
n-1}(0)}}{\rho_{n-1}(\tilde{v}_j+\tilde{v}_i)}
\frac{R^{(n-1)}(-\tilde{v}_i-\tilde{v}_j)^{c_1\bar{c}_1}_{e_2\bar{e}_2}
R^{(n-1)}(\tilde{v}_i-\tilde{v}_j)^{\bar{e}_2e_2}_{\bar{d}d}}
{T^{(\bar{e}_2)}(v_j)}\nonumber \\
& &\times S^{\bar{d}e_3\cdots e_m}_{h_2\cdots
h_m}(v_j;\{\check{v}_i,\check{v}_j\})S^{{d}h_2\cdots
h_m}_{d_1\cdots d_m}(v_i;\{\check{v}_i\}) \}
\tau_1(\tilde{v}_j;\{\tilde{v}_m\})^{d_1\cdots d_m}_{b_1\cdots
b_m},\label{fu33}\\
& &\hspace{-8mm}
R^{(n-1)}(\tilde{v}_j+\tilde{v}_i)^{\bar{a}_1h_1'}_{f_2d_1'}[T^{n-2}(v_i;\{\check{v}_i,\check{v}_j\})^{e_3\cdots
e_m}_{a_3'\cdots
a_m'}]_{a_1h_1'}[T^{m-2}(v_j;\{\check{v}_i,\check{v}_j\})^{a_3'\cdots
a_m'}_{c_3'\cdots c_m'}]_{f_2c_2'}\nonumber \\
& & \hspace{-2mm}\times S^{c_2'\cdots c_m'}_{d_2'\cdots
d_m'}(v_j;\{\check{v}_i,\check{v}_j\})S^{d_1'\cdots
d_m'}_{b_1\cdots
b_m}(v_i;\{\check{v}_i\})=\frac{\rho_{n-1}(\tilde{v}_i-\tilde{v}_j){\rho_{n-1}(0)}}{\rho_{n-1}(\tilde{v}_i+\tilde{v}_j)}
\nonumber \\
& &\times
\frac{R^{(n-1)}(-\tilde{v}_j-\tilde{v}_i)^{\bar{a}_1a_1}_{\bar{d}d}}{{T^{({d})}(v_i)}{T^{(\bar{a}_1)}(v_j)}}
S^{\bar{d}e_3\cdots e_m}_{a_2\cdots
a_m}(v_j;\{\check{v}_i,\check{v}_j\})S^{{d}a_2\cdots
a_m}_{g_1\cdots g_m}(v_i;\{\check{v}_i\}) ]\nonumber\\
&&\times \tau_1(\tilde{v}_i;\{\tilde{v}_m\})^{g_1\cdots
g_m}_{d_1\cdots d_m}\tau_1(\tilde{v}_j;\{\tilde{v}_m\})^{d_1\cdots
d_m}_{b_1\cdots b_m},\label{fu44}
\end{eqnarray}
where
\begin{eqnarray}
&& T^{(d_1)}(v_i)=k^+_d(v_i)R^{(n-1)}(2v_i-4\eta)^{dd_1}_{d_1d}
\end{eqnarray}
 and one  should note that all the repeated indices in eqs.(\ref{fu22},
\ref{fu33}, \ref{fu44}) sum over 1 to $2n-1$ except for  $c_1$ or
$a_1$. We now denote
\begin{eqnarray}
& &
\hspace{-6mm}|\tilde{\Psi}_{m-1}^{(2)}(u,v_i;\{v_m\})_{fd}\rangle=
\frac{\rho^{\frac{1}{2}}_{n-1}(0)\omega(v_i)}{T^{(d)}(v_i)}
\Lambda_2^{m-1}(v_i;\{\check{v}_i\}) B_f(u)\Phi_{m-1}^{e_2\cdots
e_m}(v_1,\cdots,\check{v}_i,\cdots,v_m)\nonumber\\
& & \hspace{10mm} \times S^{de_2\cdots e_m}_{d_1\cdots
d_m}(v_i;\{\check{v}_i\})
\tau_1(\tilde{v}_i;\{\tilde{v}_m\})^{d_1\cdots d_m}_{b_1\cdots
b_m} F^{b_1\cdots b_m}|0\rangle,\label{tphi2}\\[2mm]
& & \hspace{-6mm}
|\tilde{\Psi}_{m-1}^{(4)}(u,v_i;\{v_m\})_{ab}\rangle=\frac{\rho^{\frac{1}{2}}_{n-1}(0)
\omega(v_i)}{T^{(\bar{b})}(v_i)}
\Lambda_2^{m-1}(v_i;\{\check{v}_i\}) E_a(u)\Phi_{n-1}^{e_2\cdots
e_m}(v_1,\cdots,\check{v}_i,\cdots,v_m)\nonumber\\
& & \hspace{10mm}\times S^{\bar{b}e_2\cdots e_m}_{d_1\cdots
d_m}(v_i;\{\check{v}_i\})
\tau_1(\tilde{v}_i;\{\tilde{v}_m\})^{d_1\cdots d_m}_{b_1\cdots
b_m} F^{b_1\cdots b_m}|0\rangle,\label{tphi4}\\[2mm]
& &\hspace{-10mm}
|\tilde{\Psi}_{m-2}^{(5)}(u,v_i,v_j;\{v_m\})_{d_1}\rangle=
F(u)\Phi_{m-2}^{e_3\cdots e_m}(v_1,\cdots,\check{v}_i,\cdots,\check{v}_j,\cdots,v_m)\nonumber \\
& &\times S^{\bar{d}_1e_3\cdots e_m}_{d_2\cdots
d_m}(v_j;\{\check{v}_i,\check{v}_j\})S^{d_1\cdots d_m}_{b_1\cdots
b_m}(v_i;\{\check{v}_i\})\nonumber \\
& & \hspace{4mm}\times \Lambda_1^{m-1}(v_i;
\{\check{v}_i\})\Lambda_1^{m-1}(v_j;
\{\check{v}_j\})\omega_1(v_i)\omega_1(v_j)F^{b_1\cdots
b_m}|0\rangle,\label{tphi5}\\[2mm]
& &\hspace{-10mm}
|\tilde{\Psi}_{m-2}^{(6)}(u,v_i,v_j;\{v_m\})_{d_1}\rangle=
F(u)\Phi_{m-2}^{e_3\cdots e_m}(v_1,\cdots,\check{v}_i,\cdots,\check{v}_j,\cdots,v_m)\nonumber \\
& &\times S^{\bar{d}_1e_3\cdots e_m}_{d_2\cdots
d_m}(v_j;\{\check{v}_i,\check{v}_j\})S^{d_1\cdots d_m}_{c_1\cdots
c_m}(v_i;\{\check{v}_i\})\tau_1(\tilde{v}_i;\{\tilde{v}_m\})^{c_1\cdots
c_m}_{b_1\cdots b_m}\nonumber \\
& & \hspace{4mm}\times \Lambda_2^{m-1}(v_i;
\{\check{v}_i\})\omega(v_i){\rho^{\frac{1}{2}}_{n-1}(0)}\Lambda_1^{m-1}(v_j;
\{\check{v}_j\})\omega_1(v_j)F^{b_1\cdots
b_m}|0\rangle,\label{tphi6}\\[2mm]
& &\hspace{-10mm}
|\tilde{\Psi}_{m-2}^{(7)}(u,v_i,v_j;\{v_m\})_{d_1}\rangle=
F(u)\Phi_{m-2}^{e_3\cdots e_m}(v_1,\cdots,\check{v}_i,\cdots,\check{v}_j,\cdots,v_m)\nonumber \\
& &\times S^{\bar{d}_1e_3\cdots e_m}_{d_2\cdots
d_m}(v_j;\{\check{v}_i,\check{v}_j\})S^{d_1\cdots d_m}_{c_1\cdots
c_m}(v_i;\{\check{v}_i\})\tau_1(\tilde{v}_j;\{\tilde{v}_m\})^{c_1\cdots
c_m}_{b_1\cdots b_m}\nonumber \\
& & \hspace{4mm}\times \Lambda_2^{m-1}(v_j;
\{\check{v}_j\})\omega(v_j){\rho^{\frac{1}{2}}_{n-1}(0)}\Lambda_1^{m-1}(v_i;
\{\check{v}_i\})\omega_1(v_i)F^{b_1\cdots
b_m}|0\rangle,\label{tphi7}\\[2mm]
& &\hspace{-10mm}
|\tilde{\Psi}_{m-2}^{(8)}(u,v_i,v_j;\{v_m\})_{d_1}\rangle=
F(u)\Phi_{m-2}^{e_3\cdots e_m}(v_1,\cdots,\check{v}_i,\cdots,\check{v}_j,\cdots,v_m)\nonumber \\
& &\times S^{\bar{d}_1e_3\cdots e_m}_{d_2\cdots
d_m}(v_j;\{\check{v}_i,\check{v}_j\})S^{{d}_1\cdots
d_m}_{c_1\cdots
c_m}(v_i;\{\check{v}_i\})\tau_1(\tilde{v}_i;\{\tilde{v}_m\})^{c_1\cdots
c_m}_{a_1\cdots a_m}\nonumber \\
& & \times \tau_1(\tilde{v}_j;\{\tilde{v}_m\})^{a_1\cdots
a_m}_{b_1\cdots b_m}\Lambda_2^{m-1}(v_i;
\{\check{v}_i\})\Lambda_2^{m-1}(v_j;
\{\check{v}_j\})\omega(v_i)\omega(v_j){\rho_{n-1}(0)}F^{b_1\cdots
b_m}|0\rangle. \label{tphi8}
\end{eqnarray}

\noindent With the help of relation Eq.(\ref{ir1}), we can easily
change the $|{\Psi}_{m-1}^{(2)}(u,v_i;\{v_m\})_{fd}\rangle$ and
$|{\Psi}_{m-1}^{(4)}(u,v_i;\{v_m\})_{ab}\rangle$ into
$|\tilde{\Psi}_{m-1}^{(2)}(u,v_i;\{v_m\})_{fd}\rangle$ and
$|\tilde{\Psi}_{m-1}^{(4)}(u,v_i;\{v_m\})_{ab}\rangle$,
respectively. It is easy to get
\begin{eqnarray}
& & \hspace{-10mm} \delta_{\bar{d}_1e_2}
H^{x}_{1,d_1}(u,v_i,v_j)|\Psi_{m-2}^{(5)}(u,v_i,v_j;\{v_m\})_{d_1e_2}\rangle
=\nonumber \\
& &
\tilde{H}^{x}_{1,d_1}(u,v_i,v_j)|\tilde{\Psi}_{m-2}^{(5)}(u,v_i,v_j;\{v_m\})_{d_1}\rangle
\end{eqnarray}
with
\begin{eqnarray}
& &\hspace{-10mm} \tilde{H}^{x}_{1,d_1}(u,v_i,v_j)=
\frac{H^{x}_{1,d_1}(u,v_i,v_j)}{a^1_1(v_i,v_j)a^1_1(v_j,v_i)}.
\end{eqnarray}
Using Eq.(\ref{fu22}) and Eq.(\ref{haa2}), Eq.(\ref{fu33}),
Eq.(\ref{fu44}) and Eq.(\ref{haa4}),  we can rewrite
\begin{eqnarray}
& & \hspace{-10mm}
H^{x}_{2,d_1}(u,v_i,v_j)^{fe}_{c_2d_1}|\Psi_{m-2}^{(6)}(u,v_i,v_j;\{v_m\})_{d_1c_2}^{ef}\rangle
=\nonumber \\
& &
\tilde{H}^{x}_{2,d_1}(u,v_i,v_j)|\tilde{\Psi}_{m-2}^{(6)}(u,v_i,v_j;\{v_m\})_{d_1}\rangle,\\
& & \hspace{-10mm} H^{x}_{3,d_1}(u,v_i,v_j)
|\Psi_{m-2}^{(7)}(u,v_i,v_j;\{v_m\})_{d_1d_1}\rangle
=\nonumber \\
& &
\tilde{H}^{x}_{3,d_1}(u,v_i,v_j)|\tilde{\Psi}_{m-2}^{(7)}(u,v_i,v_j;\{v_m\})_{d_1}\rangle,\\
& & \hspace{-10mm}
H^{x}_{4,d_1}(u,v_i,v_j)^{de}_{fd_1}|\Psi_{m-2}^{(8)}(u,v_i,v_j;\{v_m\})_{d_1f}^{ed}\rangle
=\nonumber \\
& &
\tilde{H}^{x}_{4,d_1}(u,v_i,v_j)|\tilde{\Psi}_{m-2}^{(8)}(u,v_i,v_j;\{v_m\})_{d_1}\rangle,
\end{eqnarray}
respectively. Where
\begin{eqnarray}
& &\hspace{-10mm} \tilde{H}^{x}_{2,d_1}(u,v_i,v_j)=\frac{1}
{\tilde{\rho}(v_i,v_j)a^1_1(v_j,v_i)}
\frac{H^{x}_{2,{d}}(u,v_i,v_j)^{dd}_{dd}}
{R^{(n-1)}(\tilde{v}_i-\tilde{v}_j)^{dd}_{dd}}
\frac{R^{(n-1)}(-\tilde{v}_i-\tilde{v}_j)^{d\bar{d}}_{\bar{d}_1d_1}}{\rho_{n-1}(\tilde{v}_i+\tilde{v}_j)T^{(d_1)}(v_i)},\\
& &\hspace{-10mm} \tilde{H}^{x}_{3,d_1}(u,v_i,v_j)=
\frac{H^{x}_{3,{d}}(u,v_i,v_j)R^{(n-1)}(-\tilde{v}_i-\tilde{v}_j)^{d\bar{d}}_{e\bar{e}}
R^{(n-1)}(\tilde{v}_i-\tilde{v}_j)^{\bar{e}e}_{\bar{d}_1d_1}}
{\tilde{\rho}(v_j,v_i)a^1_1(v_i,v_j)\rho_{n-1}(\tilde{v}_j-\tilde{v}_i)
\rho_{n-1}(\tilde{v}_j+\tilde{v}_i)T^{(\bar{e})}(v_j)},\\
& &\hspace{-6mm} \tilde{H}^{x}_{4,d_1}(u,v_i,v_j)=\frac{1}
{\tilde{\rho}(v_i,v_j)\tilde{\rho}(v_j,v_i)\rho_{n-1}(\tilde{v}_j-\tilde{v}_i)\rho_{n-1}(\tilde{v}_i+\tilde{v}_j)}\nonumber\\
& & \hspace{25mm}\times\frac{H^{x}_{4,{d}}(u,v_i,v_j)^{dd}_{dd}}
{R^{(n-1)}(\tilde{v}_i-\tilde{v}_j)^{dd}_{dd}}
\frac{R^{(n-1)}(-\tilde{v}_j-\tilde{v}_i)^{d\bar{d}}_{\bar{d}_1d_1}}{{T^{({d_1})}(v_i)}{T^{(d)}(v_j)}}.
\end{eqnarray}
After making the  notation
\begin{eqnarray}
|\tilde{\Psi}_{x}(u,\{v_m\})\rangle=\left\{\begin{array}{l}|{\Psi}_{x}(u,\{v_m\})\rangle, x=A,\tilde{A}_2\\[2mm]
\omega(u)\Lambda_2^m(u;\{{v}_m\})\Phi_m^{d_1\cdots
d_m}(v_1,\cdots,v_m)[{T}^{m}(u;\{{v}_m\})^{d_1\cdots
d_m}_{b_1\cdots b_m}]_{aa} F^{b_1\cdots b_m}|0\rangle,
x=\tilde{A}_{aa} \end{array}\right.
\end{eqnarray}
we arrive at the final result Eq.(\ref{nd1pns}).


\section{Necessary coefficients } \setcounter{equation}{0}
\renewcommand{\theequation}{B.\arabic{equation}}

\begin{eqnarray}
&&\omega_1^{(j)}(u^{(j)};\{\tilde{v}_{m_{j-1}}^{(j-1)}\})=
\bar{\omega}_1^{(j)}(u^{(j)})\xi_1^{(j)}(u^{(j)};\{\tilde{v}_{m_{j-1}}^{(j-1)}\}),\nonumber\\
&&\omega^{(j)}(u^{(j)};\{\tilde{v}_{m_{j-1}}^{(j-1)}\})=
\bar{\omega}^{(j)}(u^{(j)})\xi_2^{(j)}(u^{(j)};\{\tilde{v}_{m_{j-1}}^{(j-1)}\}),\nonumber\\
&&\omega_{2(n-j)+1}^{(j)}(u^{(j)};\{\tilde{v}_{m_{j-1}}^{(j-1)}\})=
\bar{\omega}_{2(n-j)+1}^{(j)}(u^{(j)})\xi_3^{(j)}(u^{(j)};\{\tilde{v}_{m_{j-1}}^{(j-1)}\})
\end{eqnarray}
with
\begin{eqnarray}
&&\xi_1^{(j)}(u^{(j)};\{\tilde{v}_{m_{j-1}}^{(j-1)}\})=
\prod_{i=1}^{m_{j-1}}\frac{a_{n-j}(u^{(j)}+\tilde{v}^{(j-1)}_i)a_{n-j}(u^{(j)}-\tilde{v}^{(j-1)}_i)}
{a_{n-j}(u^{(j)}-\tilde{v}^{(j-1)}_i)a_{n-j}(\tilde{v}^{(j-1)}_i-u^{(j)})},\nonumber\\
&&\xi_2^{(j)}(u^{(j)};\{\tilde{v}_{m_{j-1}}^{(j-1)}\})=
\prod_{i=1}^{m_{j-1}}\frac{b_{n-j}(u^{(j)}+\tilde{v}^{(j-1)}_i)b_{n-j}(u^{(j)}-\tilde{v}^{(j-1)}_i)}
{a_{n-j}(u^{(j)}-\tilde{v}^{(j-1)}_i)a_{n-j}(\tilde{v}^{(j-1)}_i-u^{(j)})},\nonumber\\
&&\xi_3^{(j)}(u^{(j)};\{\tilde{v}_{m_{j-1}}^{(j-1)}\})=
\prod_{i=1}^{m_{j-1}}\frac{e_{n-j}(u^{(j)}+\tilde{v}^{(j-1)}_i)e_{n-j}(u^{(j)}-\tilde{v}^{(j-1)}_i)}
{a_{n-j}(u^{(j)}-\tilde{v}^{(j-1)}_i)a_{n-j}(\tilde{v}^{(j-1)}_i-u^{(j)})}.
\end{eqnarray}
For the case of Eq.(\ref{kn1}), we have
\begin{eqnarray}
&& \bar{\omega}_{1}^{(j)}(u^{(j)})=1,
\hspace{4mm}\bar{\omega}^{(j)}(u^{(j)})=\frac{e^{2\eta}\sinh(u^{(j)})}{\sinh(u^{(j)}-2\eta)},\nonumber
\\
&&
\bar{\omega}_{2(n-j)+1}^{(j)}(u^{(j)})=\frac{e^{2(2(n-j)-1)\eta}\sinh(u^{(j)})\cosh(u^{(j)}-(2(n-j)+3)\eta)}
{\sinh(u^{(j)}-4(n-j)\eta)\cosh(u^{(j)}-(2(n-j)+1)\eta)}.
\end{eqnarray}
while for  the case of Eq.(\ref{kn2}), if $p_-=n$
\begin{eqnarray}
&&
\bar{\omega}_{1}^{(j)}(u^{(j)})=e^{-u^{(j)}}[c_-\cosh(\eta)+\sinh(u^{(j)}-2(n-j)\eta)
], \nonumber\\
&&\bar{\omega}_{2(n-j)+1}^{(j)}(u^{(j)})=\frac{e^{u^{(j)}-4\eta}\sinh(u^{(j)})
[c_-\sinh(2\eta)+\cosh(u^{(j)}-(2(n-j)+1)\eta)]}
{\sinh(u^{(j)}-4(n-j)\eta)\cosh(u^{(j)}-(2(n-j)+1)\eta)}
\nonumber\\
&&\hspace{20mm}\times[c_-\cosh(\eta)+\sinh(u^{(j)}-2(n-j)\eta)],\nonumber\\
&&\bar{\omega}^{(j)}(u^{(j)})=\frac{\sinh(u^{(j)})}{\sinh(u^{(j)}-2\eta)}
\times\left\{\begin{array}{ll} 1,& (j\ne n-1)\\[2mm]
c_-\cosh(u^{(j)}-\eta).& (j=n-1)
\end{array}\right.
\end{eqnarray}
If $p_-\ne n$,
\begin{eqnarray}
&&
\bar{\omega}_{1}^{(j)}(u^{(j)})=e^{-u^{(j)}}[c_-\cosh(\eta)+\sinh(u^{(j)}-2(2p_--(n+j))\eta)
], \nonumber\\
&&\bar{\omega}_{2(n-j)+1}^{(j)}(u^{(j)})=\frac{e^{u^{(j)}-4\eta}\sinh(u^{(j)})
[c_-\sinh(2\eta)+\cosh(u^{(j)}-(2(n-j)+1)\eta)]}
{\sinh(u^{(j)}-4(n-j)\eta)\cosh(u^{(j)}-(2(n-j)+1)\eta)}
\nonumber\\
&&\hspace{20mm}\times[c_-\cosh((4p_--4(n-j)-1)\eta)+\sinh(u^{(j)}-2(n-j)\eta)],\nonumber\\
&&\bar{\omega}^{(j)}(u^{(j)})=\frac{\sinh(u^{(j)})}{\sinh(u^{(j)}-2\eta)}
\end{eqnarray}
and
\begin{eqnarray}
&&
\bar{\omega}_{1}^{(j)}(u^{(j)})=[c_-\cosh(u^{(j)}-(2p_--2j-1)\eta)+\sinh(2(n-p_-)\eta)], \nonumber\\
&&\bar{\omega}_{2(n-j)+1}^{(j)}(u^{(j)})=\frac{e^{2(2(n-j)-1)\eta}\sinh(u^{(j)})\cosh(u^{(j)}-(2(n-j)+3)\eta)}
{\sinh(u^{(j)}-4(n-j)\eta)\cosh(u^{(j)}-(2(n-j)+1)\eta)}
\nonumber\\
&&\hspace{20mm}\times[c_-\cosh(u^{(j)}-(2p_--2j-1)\eta)+\sinh(2(n-p_-)\eta)],\nonumber\\
&&\bar{\omega}^{(j)}(u^{(j)})=\frac{e^{2\eta}\sinh(u^{(j)})}{\sinh(u^{(j)}-2\eta)}\nonumber\\
&&\hspace{10mm}\times\left\{\begin{array}{ll} 1&(j\ne n-1) \\[2mm]
[c_-\cosh(u^{(j)}-(2p_--2j-1)\eta)+\sinh(2(n-p_-)\eta)] &
(j=n-1)\end{array}\right.
\end{eqnarray}
for $j< p_-$ and  $p_-\le j\le n-1$, respectively. For the case of
Eq.(\ref{kp1}), we have
\begin{eqnarray}
&&
w_{1}^{(j)}(u^{(j)})=\frac{\sinh(u^{(j)}-2(2(n-j)+1)\eta)\cosh(u^{(j)}-(2(n-j)-1)\eta)}
{\sinh(u^{(j)}-2\eta)\cosh(u^{(j)}-(2(n-j)+1)\eta)},\nonumber\\
&&
w^{(j)}(u^{(j)})=\frac{e^{-2\eta}\sinh(u^{(j)}-2(2(n-j)+1)\eta)}{\sinh(u^{(j)}-4(n-j)\eta)},
\hspace{4mm}w_{2(n-j)+1}^{(j)}(u^{(j)})=e^{-2(2(n-j)-1)\eta},
\end{eqnarray}
\begin{equation}
\beta_{j+1}(v_i^{(j)})=\displaystyle\left\{\begin{array}{ll}
\displaystyle-\frac{2e^{2\eta}\sinh(v_i^{(j)})\sinh(v_i^{(j)}-4(n-j)\eta)
\cosh(v_i^{(j)}-(2n-2j-1)\eta)}{\sinh(v_i^{(j)}-2\eta)},&
j\le n-2\\
\displaystyle-\frac{e^{2\eta}\sinh(v_i^{(n-1)})}{\sinh(v_i^{(n-1)}-2\eta)}.&
j=n-1
\end{array}\right.
\end{equation}
while for the case of Eq.(\ref{kp2}), if $p_+=n$
\begin{eqnarray}
&&
w_{1}^{(j)}(u^{(j)})=\frac{e^{u^{(j)}}\sinh(u^{(j)}-2(2(n-j)+1)\eta)}{\sinh(u^{(j)}-2\eta)\cosh(u^{(j)}-(2n-2j+1)\eta)}
[\sinh(2\eta)+c_+\cosh(u^{(j)}-(2(n-j)+1)\eta)]\nonumber\\
&&\hspace{20mm}\times[\cosh(\eta)-c_+\sinh(u^{(j)}-2((n-j)+1)\eta)],\nonumber\\
&&w_{2(n-j)+1}^{(j)}(u^{(j)})=K^{(2)}_{+}(u^{(j)},n-j,n-j)_{2(n-j)+1},\nonumber\\
&&w^{(j)}(u^{(j)})=
\frac{\sinh(u^{(j)}-2(2(n-j)+1)\eta)}{\sinh(u^{(j)}-4(n-j)\eta)}\times\left\{\begin{array}{ll}
1,&(j\ne n-1)\\[2mm]
c_+\cosh(u^{(j)}-5\eta).&(j= n-1)\end{array}\right.
\end{eqnarray}
If $p_+\ne n$,
\begin{eqnarray}
&&
w_{1}^{(j)}(u^{(j)})=\frac{e^{u^{(j)}}\sinh(u^{(j)}-2(2(n-j)+1)\eta)}{\sinh(u^{(j)}-2\eta)\cosh(u^{(j)}-(2n-2j+1)\eta)}
[\sinh(2\eta)+c_+\cosh(u^{(j)}-(2(n-j)+1)\eta)]\nonumber\\
&&\hspace{20mm}\times[\cosh((4p_+-4n-1)\eta)-c_+\sinh(u^{(j)}-2((n-j)+1)\eta)],\nonumber\\
&&w_{2(n-j)+1}^{(j)}(u^{(j)})=K^{(2)}_{+}(u^{(j)},n-j,p_+-j)_{2(n-j)+1},\hspace{2mm}w^{(j)}(u^{(j)})=
\frac{\sinh(u^{(j)}-2(2(n-j)+1)\eta)}{\sinh(u^{(j)}-4(n-j)\eta)}
\end{eqnarray}
and
\begin{eqnarray}
&& w_{1}^{(j)}(u^{(j)})=
\frac{\sinh(u^{(j)}-2(2(n-j)+1)\eta)\cosh(u^{(j)}-(2(n-j)-1)\eta)}
{\sinh(u^{(j)}-2\eta)\cosh(u^{(j)}-(2(n-j)+1)\eta)}\nonumber\\
&&\hspace{20mm}\times
[c_+\cosh(u^{(j)}-(4n-2p_+-2j+3)\eta)-\sinh(2(n-p_+)\eta)],\nonumber\\
&&w_{2(n-j)+1}^{(j)}(u^{(j)})=e^{-4(n-j-\frac{1}{2})\eta}[c_+\cosh(u^{(j)}-(4n-2p_+-2j+3)\eta)-\sinh(2(n-p_+)\eta)],\nonumber\\
&&w^{(j)}(u^{(j)})=\frac{e^{-2\eta}\sinh(u^{(j)}-2(2(n-j)+1)\eta)}{\sinh(u^{(j)}-4(n-j)\eta)}\nonumber\\
&&\hspace{10mm}\times\left\{\begin{array}{ll} 1,&(j\ne n-1) \\[2mm]
[c_+\cosh(u^{(j)}-(4n-2p_+-2j+3)\eta)-\sinh(2(n-p_+)\eta)] &
(j=n-1)\end{array}\right.
\end{eqnarray}
for $j< p_+$ and $p_+\le j\le n-1$, respectively,
\begin{equation}
\beta_{j+1}(v_i^{(j)})=\displaystyle \left\{\begin{array}{l}
\displaystyle
-\frac{2e^{v_i^{(j)}}\sinh(v_i^{(j)})\sinh(v_i^{(j)}-4(n-j)\eta)}{\sinh(v_i^{(j)}-2\eta)}
[\sinh(2\eta)+c_+\cosh(v_i^{(j)}-(2n-2j+1)\eta)]\nonumber\\
\hspace{4mm}\times[\cosh((4p_+-4n-1)\eta)-c_+\sinh(v_i^{(j)}-2(n-j+1)\eta)],\hspace{3mm}
(j<p_+,p_+\ne n) \\[2mm]
\displaystyle
-\frac{e^{v_i^{(n-1)}}\sinh(v_i^{(n-1)})}{\sinh(v_i^{(n-1)}-2\eta)}
[\cosh(\eta)-c_+\sinh(v_i^{(n-1)}-2\eta)],\hspace{3mm}
(p_+=n,j=n-1) \\[2mm]
\displaystyle-\frac{2e^{2\eta}\sinh(v_i^{(j)})\sinh(v_i^{(j)}-4(n-j)\eta)\cosh(v_i^{(j)}-(2n-2j-1)\eta)}{\sinh(v_i^{(j)}-2\eta)}\\[2mm]
\hspace{2mm}\times[c_+\cosh(v_i^{(j)}-(4n-2p_+-2j+3)\eta)-\sinh(2(n-p_+)\eta)],\hspace{3mm}
( p_+\le j<n-1)\\[2mm]
\displaystyle-\frac{e^{2\eta}\sinh(v_i^{(n-1)})}{\sinh(v_i^{(n-1)}-2\eta)}.\hspace{3mm}
(p_+\le j=n-1)
\end{array}\right.
\end{equation}

\end{document}